\documentclass[lettersize,journal]{IEEEtran}
\usepackage{ntheorem,amsfonts}
\usepackage{array}
\usepackage{textcomp}
\usepackage{stfloats}
\usepackage{verbatim}
\usepackage{subfigure}
\newtheorem{theorem}{Theorem}

\usepackage{lipsum}
\newtheorem{refproof}{Proof}
\usepackage{cite}
\usepackage{colortbl}
\usepackage{graphicx}
\usepackage{amsmath}
\usepackage{amssymb}
\usepackage{algorithmicx,algorithm}
\usepackage{algpseudocode}
\usepackage{booktabs}
\usepackage[pagebackref,breaklinks,colorlinks,bookmarks=false]{hyperref}
\usepackage{epsfig}
\usepackage{graphicx}
\usepackage{multirow}
\usepackage{tabularx}
\usepackage{makecell}
\usepackage{url}
\usepackage{bm}
\usepackage{threeparttable}
\usepackage{comment}
\usepackage{mathrsfs}
\usepackage{epsfig}
\usepackage{graphicx}
\usepackage{amsmath}
\usepackage{amssymb}
\definecolor{mygray}{gray}{0.8}

\title{A Proxy Attack-Free Strategy for Practically Improving the Poisoning Efficiency in Backdoor Attacks}

\author{$\text{Ziqiang Li}^{1,2} \quad \text{Hong Sun}^{2} \quad  \text{Pengfei Xia}^{3} \quad \text{Beihao Xia}^4 \quad \text{Xue Rui}^1$ \\  $ \text{Wei Zhang}^2 \quad \text{Qinglang Guo}^5 \quad \text{Zhangjie Fu}^{1,\dagger} \quad \text{Bin Li}^{2,\dagger} $ \\

$^1 \text{Nanjing University of Information Science and Technology.}$
$^2 \text{Big Data and Decision Lab, University of Science and Technology of China.}$ 
$^3 \text{Huawei Technologies Co. Ltd.}$ \\  
$^4 \text{Huazhong University of Science and Technology.}$  $^5 \text{China Academic of Electronics and Information Technology.}$  \\
{\tt\small \{iceli,hsun777,xpengfei\}@mail.ustc.edu.cn, xbh\_hust@hust.edu.cn} \\ {\tt\small \{ruixue27,zw1996,gql1993\}@mail.ustc.edu.cn, fzj@nuist.edu.cn,  binli@ustc.edu.cn}
}

\begin{document}
\maketitle
\renewcommand{\thefootnote}{\fnsymbol{footnote}}
\footnotetext[2]{Corresponding Author.}

\begin{abstract}

Poisoning efficiency is crucial in poisoning-based backdoor attacks, as attackers aim to minimize the number of poisoning samples while maximizing attack efficacy. Recent studies have sought to enhance poisoning efficiency by selecting effective samples. However, these studies typically rely on a proxy backdoor injection task to identify an efficient set of poisoning samples. This proxy attack-based approach can lead to performance degradation if the proxy attack settings differ from those of the actual victims, due to the shortcut nature of backdoor learning. Furthermore, proxy attack-based methods are extremely time-consuming, as they require numerous complete backdoor injection processes for sample selection. To address these concerns, we present a Proxy attack-Free Strategy (PFS) designed to identify efficient poisoning samples based on the similarity between clean samples and their corresponding poisoning samples, as well as the diversity of the poisoning set. The proposed PFS is motivated by the observation that selecting samples with high similarity between clean and corresponding poisoning samples results in significantly higher attack success rates compared to using samples with low similarity. Additionally, we provide theoretical foundations to explain the proposed PFS. We comprehensively evaluate the proposed strategy across various datasets, triggers, poisoning rates, architectures, and training hyperparameters. Our experimental results demonstrate that PFS enhances backdoor attack efficiency while also offering a remarkable speed advantage over previous proxy attack-based selection methodologies.

\end{abstract}


\section{Introduction}
\label{sec:intro}
The abundance of training data is crucial to the success of Deep Neural Networks (DNNs) \cite{li2023systematic}. For example, GPT-3 \cite{brown2020language}, a deep learning model with 175 billion parameters, owes its effectiveness in various language processing tasks to its pre-training on a massive corpus of 45 TB of text data. In response to the increasing demand for data, users and businesses are increasingly turning to third-party sources or online repositories for more convenient data collection. However, recent researches \cite{chen2017targeted,gu2019badnets,goldblum2022dataset,biggio2012poisoning,cina2023wild,biggio2018wild,li2022new} have demonstrated that these practices can be maliciously exploited by attackers to poison training data, thereby negatively impacting the functionality and reliability of trained models.


During the training phase, a significant threat emerges in the form of \textit{backdoor attacks} \cite{chen2017targeted,gu2019badnets,liu2017trojaning,li2024efficient}. These attacks involve injecting a covert backdoor into DNNs by introducing a small number of poisoned samples into an a benign training dataset. While the model appears normal when presented with benign inputs, a predefined trigger can activate the infected model, forcing its predictions to align with the attackers' objectives. As research continues to reveal this threat in various tasks, such as speaker verification \cite{zhai2021backdoor}, malware detection \cite{li2021backdoor}, emotion analysis \cite{zeng2023efficient,li2024largelanguagemodelsgood}, and Deepfake detection \cite{sun2023real}, attention to backdoor attacks is increasing.

Recent studies have sought to enhance the efficiency of backdoor attacks by selecting optimal poisoning samples. These studies \cite{xia2022data,gao2023not,guo2023temporal,li2023explore} demonstrate that an equivalent attack success rate can be achieved using only 47\% to 75\% of the poisoning samples compared to random selection strategies. However, it has become apparent that these methods rely on a proxy backdoor task to assess the contribution of each poisoning sample, thus constructing an efficient poison set closely associated with the specific proxy attack of the attacker. These proxy attack-based selection methods \cite{xia2022data,gao2023not,guo2023temporal} pose a significant challenge due to potential deterioration in its effectiveness when the proxy attack settings of attackers deviate from the actual settings of the victims—a frequent scenario in attacks \cite{zhao2020clean,zhong2020backdoor,zeng2022narcissus}. Moreover, conducting the proxy attack task significantly increases the execution time of the algorithm, limiting its scalability on large datasets.

\begin{figure*}[t!]
    \setlength{\abovecaptionskip}{0.1cm}
    \setlength{\belowcaptionskip}{-0.3cm}
    \vspace{-1em}
	\centering
	\includegraphics[scale=0.4]{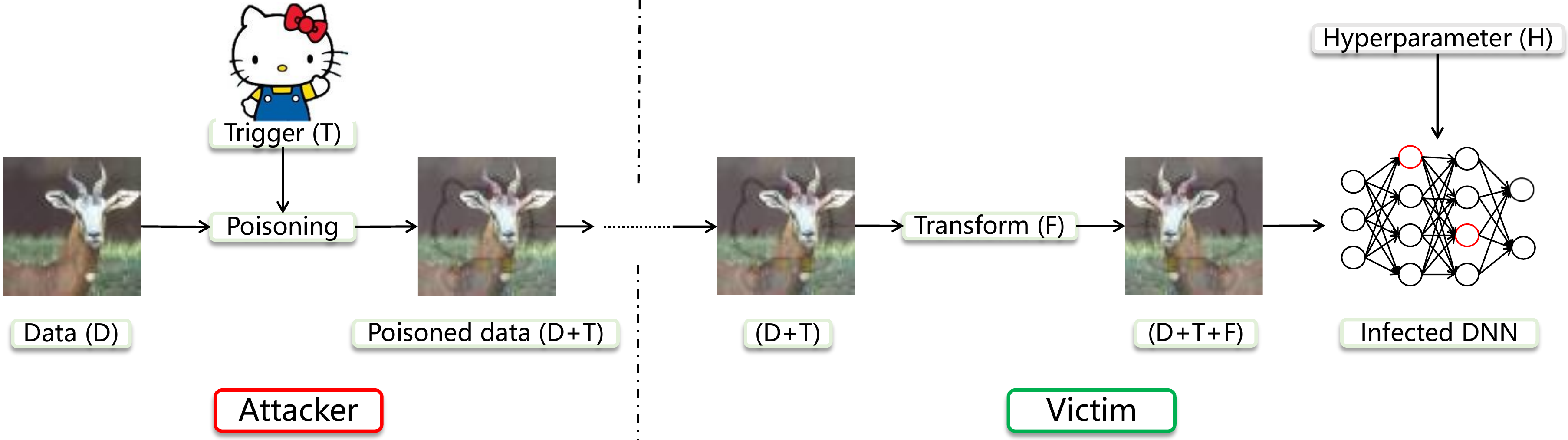}
	\caption{The pipeline of poisoning-based backdoor attacks typically involves an attacker who combines a clean dataset (D) with a trigger (T) to create a poisoned dataset (D+T), which is then released to the victims. The victims download the poisoned data and use it to train their DNN models, applying various data transformations and augmentations (F) and hyperparameters (H)\textsuperscript{\ref{foot2}} during training. As a result, the DNN models can be infected with the backdoor trigger, which can be activated by a specific trigger condition during the inference phase.}
	\label{fig:data-efficient}
\end{figure*}

This paper addresses the aforementioned challenges by introducing a novel \textit{Proxy attack-Free Strategy} (PFS). Unlike previous methods reliant on proxy attack tasks, PFS leverages similarity measures between benign and corresponding poisoning samples. Specifically, we utilize a pre-trained feature extractor to compute cosine similarity within the feature space. We demonstrate that selecting samples with high similarity between benign and corresponding poisoning samples significantly enhances attack success rates compared to low-similarity samples. However, solely using high-similarity samples for poisoning resulted in lower success rates compared to random sampling. We attribute this performance decline to the overly constrained diversity within the poisoning sample pool. Therefore, our paper emphasizes both similarity between benign and corresponding poisoning samples and diversity within poisoning samples as critical factors influencing sample efficiency in backdoor attacks. Additionally, we provide theoretical foundations to elucidate this phenomenon. Based on these insights, we propose PFS, which incorporates a pre-trained feature extractor to filter out inefficient samples based on cosine similarity and integrates random selection to enhance sample diversity within the poisoning set. We evaluate this strategy across CIFAR-10, CIFAR-100, and Tiny-ImageNet datasets. Our experimental results demonstrate that PFS significantly improves backdoor attack efficiency while offering a notable speed advantage over proxy attack-based methodologies. Contributions can be summarized as follows:

\begin{itemize}
\item We conduct a comprehensive study of the forensic properties inherent in efficient poisoning samples within the context of backdoor attacks. 
\item Based on the insights from our analysis, we introduce a novel selection approach named the Proxy attack-Free Strategy (PFS). Unlike previous studies, PFS eliminates the requirement for a proxy attack task, effectively addressing challenges arising from differing settings and significantly reducing the time overhead.
\item The proposed PFS has achieved superior attack success rates while achieving a remarkable acceleration when contrasted with prior proxy attack-based selection methodologies. This attribute renders PFS particularly well-suited for scalability, making it apt for accommodating larger datasets and complex models.
\end{itemize}

\section{Related Works}

\subsection{Backdoor Attacks}


Backdoor attacks aim to inject Trojans within DNNs. This manipulation empowers poisoning models to yield accurate outcomes when presented with clean samples, while displaying anomalous behavior with samples containing designed triggers. These attacks can occur at various stages within the DNN development lifecycle, including code \cite{bagdasaryan2021blind,ramakrishnan2020backdoors}, outsourcing \cite{xia2022enhancing}, pre-trained models \cite{wang2020backdoor,ge2021anti}, data collection \cite{gu2019badnets,chen2017targeted,turner2019label,liao2018backdoor}, collaborative learning \cite{xu2019cryptonn,nguyen2020poisoning}, and even post-deployment scenarios \cite{rakin2020tbt,hong2019terminal}. Of these ways, the most straightforward and widely adopted approach is poisoning-based backdoor attacks, entailing the insertion of backdoor triggers through modifications to the training data. Recent researches on poisoning-based backdoor attacks have been proposed to enhancing poisoning efficiency from two aspects.

\noindent\textbf{Designing Efficient Triggers.}
Trigger design is a hot topic in backdoor learning. Chen \textit{et al.} \cite{chen2017targeted} initially proposed a blended strategy for evading human detection, acquiring poisoning samples by blending clean samples with triggers. Subsequent studies have explored leveraging natural patterns such as warping \cite{nguyen2021wanet}, rotation \cite{wu2022just,xu2023batt}, style transfer \cite{cheng2021deep}, frequency \cite{feng2022fiba,zeng2021rethinking}, and reflection \cite{liu2020reflection} to create triggers that are more imperceptible and efficient compared to previous methods. Drawing inspiration from Universal Adversarial Perturbations (UAPs) \cite{moosavi2017universal} in adversarial examples, some research \cite{zhong2020backdoor,li2020invisible,doan2021backdoor} optimizes an UAP on a pre-trained clean model as the trigger, a highly effective and widely adopted approach. 
Unlike previous approaches that use universal triggers, Li \textit{et al.} \cite{li2021invisible} employ GANs models to generate sample-specific triggers.

\noindent\textbf{Selecting Efficient Poisoning Samples.}
Efficient poisoning samples selection is an under-explored area and is orthogonal to the trigger design. Xia \textit{et al.} \cite{xia2022data} first explored the contribution to backdoor injection for different data. Their work illustrated the inequitable influence of individual poisoning samples and underscored the enormous potential for improved data efficiency within backdoor attacks through efficient sample selection. Simultaneously, Gao \textit{et al.} \cite{gao2023not} assert a similar notion—that not all samples equally foster the process of poisoning in backdoor attacks. They introduce the concept of "forgetting events" to spotlight the most potent poisoning samples in the context of clean-label backdoor attacks. Additionally, the RD score \cite{wu2023computation} measures the distance between the model’s output and the target class, enabling the selection of the most impactful samples for poisoning. To further refine this process, the authors employ a greedy search algorithm that iteratively selects samples with the highest RD scores. However, all of these studies \cite{xia2022data,gao2023not,wu2023computation} rely on proxy attacks to determine the efficient subset of samples. Regrettably, this dependence on proxy attacks introduces a vulnerability, leading to a degradation in attack performance when disparities arise in parameters like Transformation (F) and Hyperparameters (H) between the proxy poisoning attack and the actual poisoning process, as depicted in \figurename~\ref{fig:data-efficient}. Therefore, a sample selection strategy that is not based on proxy attacks is urgently needed, and is the main contribution of our paper. Comparable to our approach, the Outlier Poisoning Strategy (OPS) \cite{guo2023temporal} also utilizes a surrogate model for efficient sample selection in backdoor attacks against anti-spoof rebroadcast detection. The primary concept behind this strategy involves embedding triggers into samples that pose the greatest classification challenge (referred to as outlier samples), encouraging the network to depend on the trigger for classification. However, OPS \cite{guo2023temporal} is specifically designed for binary classification and clean-label settings, exhibiting diminished performance in scenarios involving dirty-label settings and multiple classifications. 

Recently, a few works also exploited backdoor for positive purposes \cite{li2022untargeted,li2023black,guo2023domain,ya2024towards}, which are out of our scope.

\subsection{Backdoor Defenses}

Backdoor attacks pose a significant threat to the security of Deep Neural Networks (DNNs), leading to the emergence of various defense mechanisms aimed at mitigating these risks. The current landscape of backdoor defenses can be broadly categorized into two classes: backdoor detection and backdoor erasure. Backdoor detection aims to identify the presence of poisoning either in input data or within the employed DNNs themselves \cite{gao2019strip,hayase2021spectre,dong2021black,xiang2022post,hou2024ibd,li2024nearest}. Conversely, backdoor erasure strategies aim to repair poisoning models by removing the embedded backdoor, either during or after the training process. In the training phase, the concept of training clean models on poisoned data has been introduced by \cite{li2021anti}. Post-training, a range of techniques, including fine-tuning, distillation, and pruning, can be employed to eliminate the backdoor from the poisoning model \cite{li2021neural,wu2021adversarial}.






\begin{figure*}[t!]
\vspace{-1em}
    \setlength{\abovecaptionskip}{0.1cm}
    \setlength{\belowcaptionskip}{-0.3cm}
	\centering
	\includegraphics[scale=0.48]{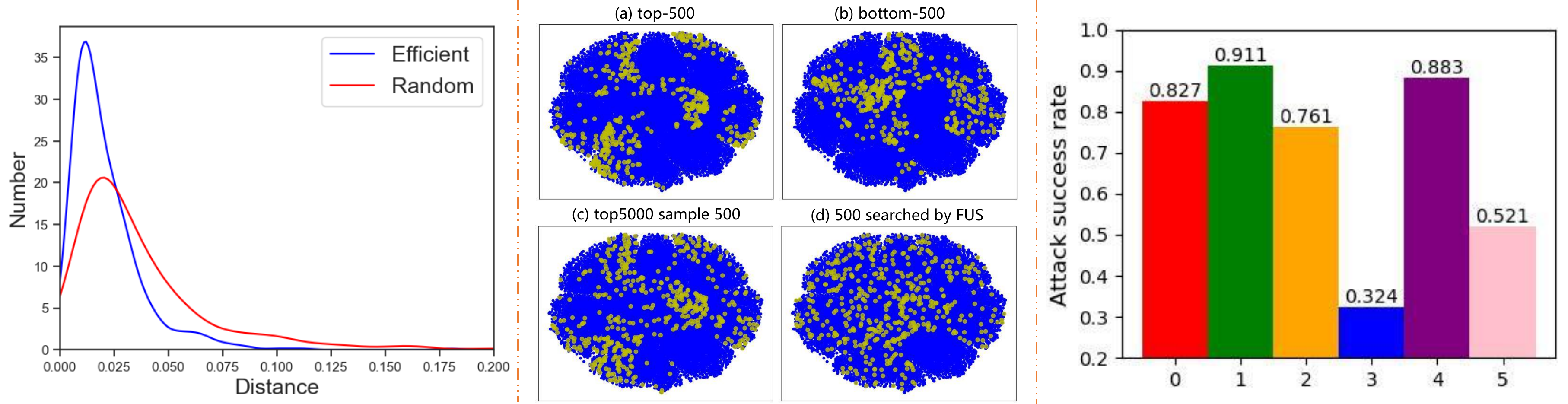}
	\caption{Three visualizations of the similarity distribution and ASR using different poisoning samples. On the \textbf{left}, the cosine similarity between benign and corresponding poisoning samples within the feature space of a pre-trained ResNet model is illustrated. This visualization encompasses a set of 500 samples obtained through random sampling and a set of 500 samples obtained through the FUS-search process. The horizontal axis corresponds to the cosine distance. On the \textbf{middle} part, we present a t-SNE visualization that contains the \textcolor[RGB]{0,0,255}{complete 50000-sample dataset} and a \textcolor[RGB]{187,185,5}{subset of 500 samples} selected using different similarity-based sampling methods. This includes four sets: Top-500 similarity samples, Bottom-500 similarity samples, Top-5,000 similarity random 500 sampling, and FUS-searched 500 samples. Lastly, on the \textbf{right} part, we provide an illustration of the ASR attained using different sets of 500 poisoning samples. The horizontal axis is annotated with labels '0', '1', '2', '3', '4', and '5', representing sifferent sampling methods: \textcolor[RGB]{255,0,0}{Random sampling}, \textcolor[RGB]{0,150,1}{FUS-selected}, \textcolor[RGB]{255,165,0}{Top-500 similarity sampling}, \textcolor[RGB]{0,0,255}{Bottom-500 similarity sampling}, \textcolor[RGB]{129,0,127}{Top-5K similarity random 500 sampling}, and \textcolor[RGB]{255,191,205}{Bottom-5K similarity random 500 sampling}, respectively. Each displayed result is the average outcome derived from five different runs.}
	\label{fig:data}
\end{figure*}

\section{Methodology}
To improve poisoning efficiency, it is important to develop proxy attack-free sample selection methods that leverage the forensic features of efficient data in poisoning-based backdoor attacks. Before introducing our method, we first review the pipeline of backdoor attacks and discuss some observations.
\subsection{Poisoning-based Backdoor Attacks.}
Poisoning-based backdoor attacks involve injecting a backdoor into a subset $\mathcal{P'}$ of a clean training set $\mathcal{D}=\{(x_i,y_i)|i=1,\cdots,N\}$, resulting in a corresponding poisoned set $\mathcal{P}=\{(x'_i,k)|x'_i=\mathcal{T}(x_i,t), (x_i,y_i) \in \mathcal{P^{'}}, i=1,\cdots,P\}$. Here, $\mathcal{T}(x,t)$ is a pre-defined generator that adds the trigger $t$ into clean sample $x$, $y_i$ and $k$ represent the true and attack-target labels of the clean sample $x_i$ and the poisoning sample $x'_i$, respectively. After poisoning, victim will train a deep model:

\begin{equation}
\begin{aligned}
\underset{\theta}\min \quad \frac{1}{N} \sum_{(x, y) \in \mathcal{D}} L\left(f_\theta(x), y\right)+
\frac{1}{P} \sum_{\left(x^{\prime}, k\right) \in \mathcal{P}} L\left(f_\theta\left(x^{\prime}\right), k\right)
\end{aligned}\text{.}
\end{equation}
In the above equation, $f_{\theta}$ represents the DNN model, and $L$ is the corresponding loss function. The poisoning rate $r$ can be calculated as $r=\frac{P}{N}$. The goal of sample selection methods is to enhance the efficiency of backdoor attacks by selecting an optimal poisoning set $\mathcal{P}$.

\subsection{Forensic Features of Efficient Data in Poisoning-based Backdoor Attacks}
\label{sec:Forensic}

We assert that the primary reason for the efficiency of backdoor injection is due to the similarity between benign and corresponding poisoning samples. Typically, poisoning samples exhibit different labels than their clean counterparts, signifying that these pairs share similar attributes yet are categorized differently in dirty-label backdoor attacks. In this context, samples exhibiting a high degree of similarity can be considered challenging samples in the context of poisoning tasks. Consequently, they tend to be more efficient than low-similarity samples for the purpose of backdoor attacks. To substantiate this assumption, we conducted empirical investigations utilizing the CIFAR-10 dataset while adhering to the same experimental settings as detailed in Sec. \ref{sec:exp_setting}. The left segment of Figure \ref{fig:data} visually captures the distribution of similarities among 500 randomly chosen samples and 500 samples searched using the FUS method \cite{xia2022data}. The result demonstrates that the FUS-searched efficient sample exhibits higher similarity compared to randomly selected samples.

We proceeded to explore the contribution of similarity between benign and corresponding poisoning samples to the efficacy of backdoor attacks. The right part of Figure \ref{fig:data} showcases the attack success rate across varying sets of poisoning samples. Notably, our observations align with our expectations: using high-similarity samples (\textcolor[RGB]{255,165,0}{Top-500 similarity sampling}) for poisoning yielded significantly higher attack success rates than utilizing low-similarity samples (\textcolor[RGB]{0,0,255}{Bottom-500 similarity sampling}). Furthermore, we compared random sampling with similarity-based selection. However, both high-similarity samples (\textcolor[RGB]{255,165,0}{Top-500 similarity sampling}) and low-similarity samples (\textcolor[RGB]{0,0,255}{Bottom-500 similarity sampling}) for poisoning yielded lower attack success rates compared to random sampling (\textcolor[RGB]{255,0,0}{Random sampling}). This implies that while similarity can effectively filter out inefficient samples, it lacks a positive correlation with data efficiency in the context of poisoning attacks.

We argue that the degeneration observed in the performance of the high-similarity samples (\textcolor[RGB]{255,165,0}{Top-500 similarity sampling}) can be attributed to the excessively constrained diversity within the pool of poisoning samples. The middle part of Figure \ref{fig:data} further illustrates this point by displaying the t-SNE representation of the entire 50,000-sample dataset and the 500 selected samples. It is evident that the high-similarity samples exhibit limited diversity. To address this limitation, we sampled 500 specimens randomly from the top 5000 samples with the highest similarity (\textcolor[RGB]{129,0,127}{Top-5K similarity random 500 sampling}). This alleviates the diversity constraint, as confirmed by the right segment of Figure \ref{fig:data}. Among these approaches, the \textcolor[RGB]{129,0,127}{Top-5K similarity random 500 sampling} method is more efficient than random sampling method.


In summary, the similarity between benign and corresponding poisoning samples and the diversity within the poisoning sample set emerge as two pivotal factors influencing the sample efficiency in backdoor attacks.

\subsection{Threat Model}

\noindent\textbf{Attacker's Capacities.}
In accordance with the fundamental prerequisites of poisoning-based backdoor attacks \cite{li2022backdoor,xia2022data}, we assume that attackers possess the ability to poison a fraction of the training dataset. However, they do not possess access to any additional training components encompassed within the victim phase, including the training loss, schedule, or model architecture. This assumption describes a more challenging and realistic scenario \cite{li2022backdoor,wu2024backdoorbench}, illustrating the attackers' constrained knowledge about the target system. Furthermore, we also assume  that attackers require a pre-trained feature extractor derived from the accessed training dataset. This assumption is widely acknowledged within the current backdoor attacks \cite{zhong2020backdoor,li2022backdoor}.


\noindent\textbf{Attacker's Goals.}
Our objective is to augment the effectiveness of dirty-label poisoning backdoor attacks costless, achieved through the judicious selection of a suitable poisoning set $\mathcal{P}$ from the dataset $\mathcal{D}$. This approach is orthogonal to  most of the current methods for improving poisoning efficiency through designing trigger. Therefore, our method can be integrated to other attacks at a fraction of the cost.

\begin{algorithm}[t]
\caption{Proxy attack-Free Strategy (PFS)} 
\label{algorithm:1}
\hspace*{0.08in} {\bf Input:}
  Clean training set $\mathcal{D}$; Size of clean training set $N$;  Backdoor trigger $t$; Attack target $k$; poisoning rate $r$; Diversity rate $m$; Pre-trained feature extractor $E$; Trigger function $\mathcal{T}$      \\
\hspace*{0.08in} {\bf Output:}
Build the poisoned set $\mathcal{P}$;
\begin{algorithmic}[1]
\State Initializing the similarity set $S$ with $\{\}$;
\For{i=1, 2, $\cdots$, N} 
\State Given a clean data $x_i$ from $\mathcal{D}$ ;
\State Adding trigger into $x_i$ to obtain poisoned data $x'_i=\mathcal{T}(x_i,t)$;
\State Computing the similarity between benign and corresponding poisoning samples in feature space $s_i=cos(E(x_i),E(x'_i))$;
\State Adding $s_i$ into $S$;
\EndFor
\State Selecting  most similar $m \times r$ samples according to $s_i$ from $\mathcal{D}$ and forming the coarse poisoned set $\mathcal{\hat{P}}$;
\State Randomly sampling $r$ samples from $\mathcal{\hat{P}}$ to form the poisoned set $\mathcal{{P}}$;

\State \Return the poisoned set $\mathcal{{P}}$
\end{algorithmic}
\vspace{-0.4em}
\end{algorithm}

\subsection{Proxy attack-Free Strategy}
Drawing inspiration from our analysis of forensic features contributing to efficient data in backdoor attacks, as showcased in Sec. \ref{sec:Forensic}, we introduce a simple yet efficient sample selection strategy termed the PFS to improve poisoning efficiency. Unlike prior approaches that rely on a proxy attack task for sample selection, our technique leverages the similarity between benign and corresponding poisoned samples, which depends solely on the Data (D) and Trigger (T), freeing it from dependence on the settings of the actual attack during the victim phase. We adopt a pre-trained feature extractor $E(\cdot)$ to compute the cosine similarity $cos(\cdot)$ within the feature space between benign and corresponding poisoning samples. Specifically, for a given benign image $x_i \in\mathcal{D}$, Backdoor trigger $t$, and Trigger function $\mathcal{T}$,  the corresponding poisoned image is defined as $x'_i=\mathcal{T}(x_i,t)$. Consequently, the similarity between benign and corresponding poisoned samples in the feature space can be computed by:

\begin{equation}
\begin{aligned}
s_i=cos(E(x_i),E(x'_i))
\end{aligned}\text{.}
\end{equation}
The similarity set is denoted as $S=\{s_1,\cdots,s_i,\cdots,s_N\}$, where $N$ is the size of the benign training set, which contains the top 
$m \times r$ most similar samples from the benign training set 
$\mathcal{D}$ based on their similarity scores. This step filters out inefficient samples. Then, to ensure diversity, we randomly select 
$r$ samples from the coarse poisoning set $\mathcal{\hat{P}}$ to form the final poisoning set, where $m$ is a hyper-parameter that modulates the diversity of the poisoning samples.

In summary, the procedure of our methodology is presented in Algorithm \ref{algorithm:1}. While PFS relies on a feature extractor pre-trained on a clean dataset to compute similarity, we argue that the essence of FUS and PFS diverges significantly. PFS identifies the forensic features of efficient samples in poisoning-based backdoor attacks and uses similarity and diversity to select efficient samples, independent of the actual attack within the victim phase. Therefore, the effectiveness of PFS primarily depends on the accuracy of similarity measurement.


Furthermore, we confirm that our method efficiently filters out most of the ineffective samples, thus enhancing the performance of FUS \cite{xia2022data}. Therefore, we introduce a collaborative strategy called FUS+PFS, which combines FUS search with the initial coarse poison set $\mathcal{\hat{P}}$ containing $m \times r$ samples to further enhance poisoning efficiency.

\subsection{Theoretical Analyses}
Additionally, we provide theoretical analyses of the proposed PFS in terms of both active learning (Sec. A of the Supplementary Materials.) and neural tangent kernel.

\noindent\textbf{Perspective of Neural Tangent Kernel (NTK).}
To explain the interesting phenomenon (\textit{i.e.},  similarity between benign and corresponding poisoning samples and the diversity within the poisoning sample set emerge as two pivotal factors influencing the efficiency of data in backdoor attacks), we exploit recent studies on NTK \cite{jacot2018neural,guo2021aeva,guo2023scale,li2023towards} for analyzing the sample efficiency of backdoor attacks: 
\begin{theorem} \label{thm:neat}
Suppose the training dataset consists of $N$ benign samples $\{(x_i,y_i)\}^{N}_{i=1}$ and $P$ poisoned samples $\{(x'_i,k)\}^{P}_{i=1}$, whose images are i.i.d. sampled from uniform distribution and belonging to $m$ classes. Assume that the DNN $f_\theta(\cdot)$ is a multivariate kernel regression $K(\cdot)$ and is trained via $\underset{\theta}\min \quad  \sum_{i=1}^{N} L\left(f_\theta(x), y\right)+
 \sum_{i=1}^{P} L\left(f_\theta\left(x^{\prime}\right), k\right)$. For the expected predictive confidences over the target label $k$, we have: $\mathbb{E}_{{x'_t}}\left[f_\theta({x'_t})\right]\propto \sum_{i=1}^{P} \frac{1}{({x}'_{{i}}-{x}_{{i}})\cdot ({x}'_{{i}}-{x}'_{{t}})}, i=1,\cdots,P $, where ${x'_t}$ is poisoning testing samples of attacks.
\end{theorem}


Theorem \ref{thm:neat} indicates that the confidence in predicting poisoned samples to the target class, $\mathbb{E}_{{x'_t}}\left[f_\theta({x'_t})\right]$ is inversely proportional to the values of ${x}'_{{i}}-{x}'_{{t}}$ and ${x}'_{{i}}-{x}_{{i}}$.  This suggests that selecting samples characterized by high similarity between benign ($x_i$) and corresponding poisoning samples ($x'_i$), as well as high similarity between poisoning samples ($x'_i$) and poisoned testing samples ($x'_t$), enhances the confidence in predicting poisoned samples to the target class. In the context of poisoning-based backdoor attacks, where training samples share the same distribution as testing samples, high similarity between poisoning samples and poisoned testing samples indicates congruence in the distribution between the poisoning set and the training set, signifying a high level of diversity within the poisoning set. Therefore, this theorem indicates that \textit{the similarity between benign and corresponding poisoning samples and the diversity within the poisoning sample set are pivotal factors influencing the efficiency of data in backdoor attacks.}  The proof of this theorem can be found in the Sec. B of the \textbf{Supplementary Materials}.

\section{Degradation of proxy attack-based Selection}
\label{sec:section3}

This section demonstrates that proxy attack-based sample selection methods experience significant degradation in sample efficiency when there is a substantial difference between the settings used for proxy poisoning attacks during the attacker phase and those employed in the actual poisoning process during the victim phase. The purpose of this section is to highlight the limitations of sample selection methods that rely on proxy attacks, thereby motivating the design of a proxy attack-free sample selection strategy for backdoor attacks. To better understand our key observations, we will first review the sample selection methods based on proxy attacks.



\noindent\textbf{Proxy Attacks-based Sample Selection.}
Recent investigations \cite{xia2022data,yang2022not,gao2023not} have highlighted that diverse samples contribute differently to the backdoor injection and a particular subset with efficient data has higher attack success rate. Notably, \cite{xia2022data,gao2023not} claim that the efficacy of poisoning samples is intertwined with instances of "forgetting events" during the backdoor injection process. 
Although FUS \cite{xia2022data}, Gradient \cite{gao2023not}, and OPS \cite{guo2023temporal} have acquired remarkable performance, they use a proxy attack processing to find the efficient subset.

As illustrated in \figurename~\ref{fig:data-efficient}, poisoning-based backdoor attacks contain two phases: \underline{Attacker} combines data (D) and trigger (T) to build poisoned data (D+T) and release it. \underline{Victims} download the released data and adopt it to train infected DNNs. Notably, during the training process, various data transformations/augmentations (F) and hyperparameters (H)\footnote{In this context, H encompasses various training settings, encompassing elements like the model architecture, optimizer choice, and training hyperparameters. \label{foot2}} have been applied by victims. Generally, attacker has no access to any information of the victim phase and only performs the poisoning injection at the attack phase. However, proxy attack-based sample selection methods need to select samples with the help of the complete attack process (both attacker and victim phases).  We argue that all parts involved in backdoor attacks are the factors that affect the efficiency of poisoning samples, containing D, T, F, and H. When there contains some gaps (\textit{e.g.} Transform (F) and Hyperparameters (H) in \figurename~\ref{fig:data-efficient}) between proxy poisoning attack and actual poisoning process of victims, the poisoning effectiveness of the selected samples may be decreased due to the shortcut nature of backdoor learning. Following, we will explore what will happen if the actual poisoning process is different from the proxy poisoning attack in proxy attack-based sample selection methods.

\subsection{Experimental Settings}
\label{sec:exp_setting}

To investigate the proposed question within this section, we engage in a series of experiments utilizing the VGG16 model \cite{simonyan2014very} as the victim model. These experiments are conducted on the CIFAR-10 and Tiny-ImageNet datasets. Our chosen config encompasses the SGD optimizer, initially set with a learning rate of 0.01, a momentum value of 0.9, and a weight decay coefficient of 5e-4. Notably, we incorporate two widely adopted data transformations—random cropping and random horizontal flipping. The training is performed for 70 epochs with a batch size of 256. Poisoning samples $x'_i$ are generated using a blending strategy \cite{chen2017targeted} in which the clean samples $x_i$ are blended with a trigger image $t$ using the equation $x'_i=\mathcal{T}(x_i,t)={{\lambda}{\cdot}{t}}+(1-{\lambda}){\cdot}{x_i}$, where ${\lambda}$ is set to 0.15. We set category 0 as the target for the attack, and we set the poisoning rate with $r=0.01$—equating to a poisoning set size of $P=500$. The process involves 15 iterations of selection, aligning with the settings of \cite{xia2022data}. In accordance with previous literature \cite{li2022backdoor}, the attack success rate is defined as the probability of classifying a poisoned test data to the target label. Unless stated otherwise, all experiments in this section follow the aforementioned settings. To explore the degeneration in poisoning efficiency within selected samples due to disparities between the proxy poisoning attack and the actual poisoning process of victims, we compare the Filtering-and-Updating Strategy (FUS) with SELF-FUS. SELF-FUS represents an ideal scenario for FUS, assuming the attacker possesses complete information about both the attacker and victim phases during poisoning-based backdoor attacks—a scenario that doesn't align with the threat model in this study. Therefore, proxy poisoning attack of attacker phase is consistent with the actual poisoning process of victim phase in SELF-FUS, which is not guaranteed in the FUS\footnote{Similarly, for Gradient \cite{gao2023not} and OPS \cite{guo2023temporal}, we compare the performance of these methods with their respective idealized versions, SELF-Gradient and SELF-OPS. SELF-Gradient and SELF-OPS represent scenarios where the attacker has complete information about both the attack and victim phases during poisoning-based backdoor attacks.}.

\begin{figure}[t!]
\setlength{\abovecaptionskip}{0.1cm}
    \setlength{\belowcaptionskip}{-0.3cm}
    \vspace{-1em}
    \centering
    \includegraphics[width=0.45\textwidth]{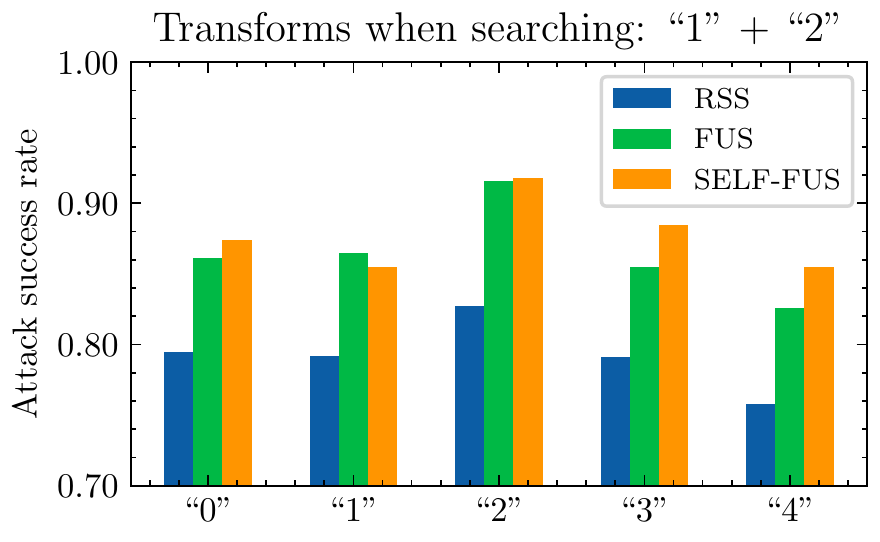}
    \caption{The ASR in scenarios where the data transformations (F) used during the proxy poisoning attack within attacker phase in FUS is different from the actual poisoning process within victim phase. SELF-FUS represents an ideal scenario for FUS, assuming the proxy poisoning attack within attacker phase is consistent with the actual poisoning process within victim phase in SELF-FUS, which is not guaranteed in the FUS.  The horizontal axis is annotated with labels "0", "1", "2", "3", and "4", corresponding to different data transformations: "None", "RandomCrop", "RandomHorizontalFlip", "RandomRotation", and "ColorJitter", respectively. Each reported result is an average computed from five separate runs.}
    \label{motivation_transform}
\end{figure}

\subsection{Empirical Studies}

\noindent\textbf{Impact of Different Data Transformations (F) on Proxy Attack-based Sample Selection Efficacy.}
We investigated how F affect the effectiveness of proxy attack-based sample selection methods. To conduct our analysis, we used the experimental settings described previously and followed them exactly. Specifically, we configured the data transformations for the proxy poisoning attack during the sample selection phase (FUS) as "RandomCrop" and "RandomHorizontalFlip". Subsequently, during the actual poisoning attack within the victim phase, we introduced variations by employing a range of transformations, including "None", "RandomCrop", "RandomHorizontalFlip", "RandomRotation", and "ColorJitter". Our results, depicted in Figure \ref{motivation_transform}, indicate that different data transforms (F) decrease the effectiveness of proxy attack-based sample selection methods. An exception to this pattern emerges in cases where the settings for both the proxy and actual attacks align closely, exemplified by the "1+2" searching and "1" attacking scenario. This particular outcome is logical and can be attributed to the inherent margin of error. 

\begin{figure}[t!]
    \vspace{-1em}
    \centering
    \includegraphics[width=0.45\textwidth]{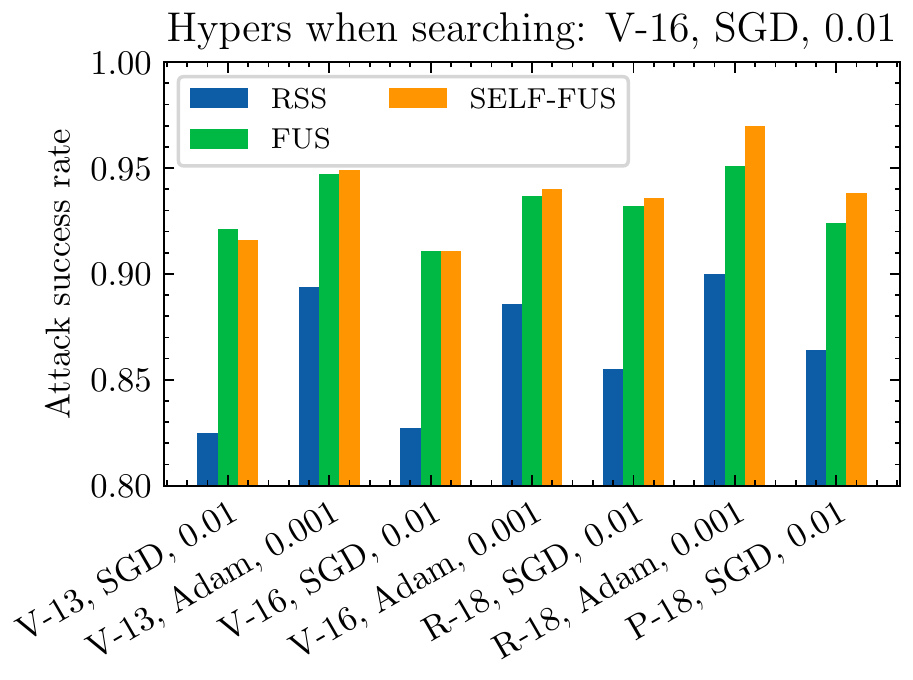}
    \caption{The ASR in scenarios where the hyperparameters (H) used during the proxy poisoning attack within attacker phase in FUS is different from the actual poisoning process within victim phase. The coordinates "V-13, SGD, 0.01"\textsuperscript{\ref{foot1}} on the horizontal axis symbolize the utilization of VGG-13 as the architecture and SGD with an initial learning rate of 0.1 as the optimizer. Each reported result is an average computed from five separate runs.}
    \label{motivation_hyper}
\end{figure}

\begin{figure*}[t!]
    \centering
    \subfigure[]{\includegraphics[width=0.31\textwidth]{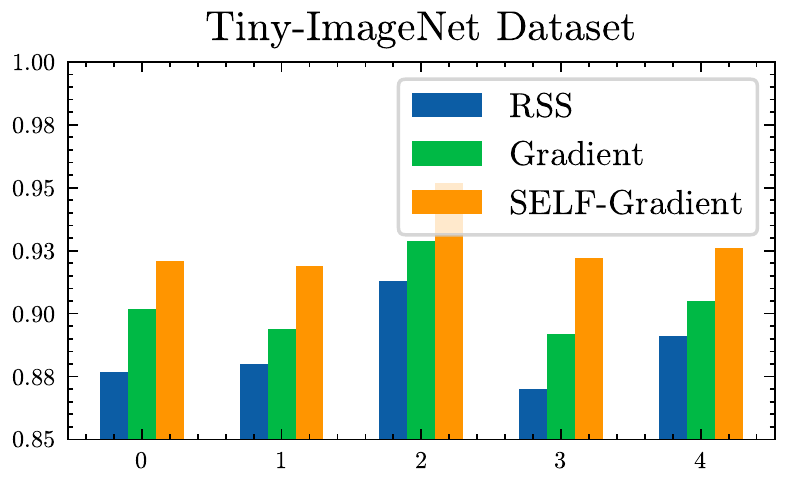}
\label{fig:SELF_GRADIENT}}
\subfigure[]{\includegraphics[width=0.31\textwidth]{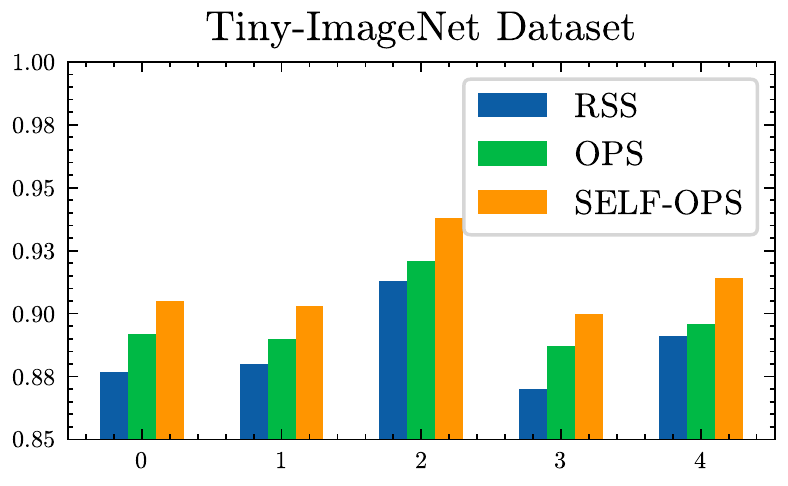} \label{fig:self_ops}}
\subfigure[]{\includegraphics[width=0.31\textwidth]{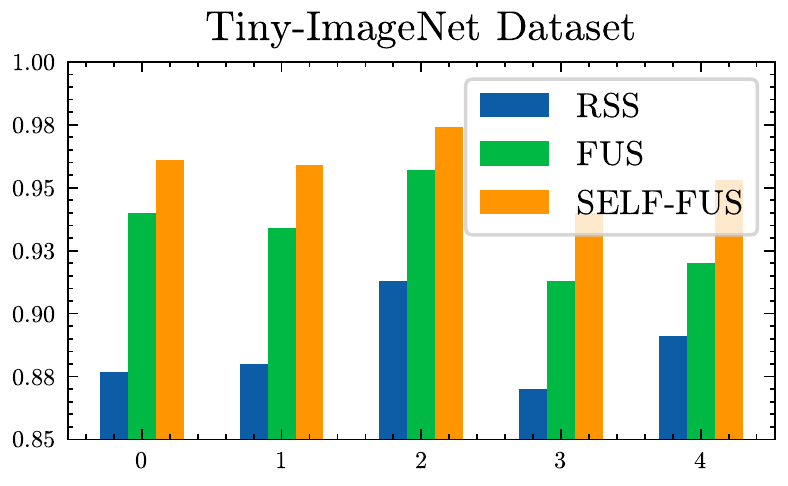} \label{fig:self_FUS}}
    \caption{The attack success rate in situations where both the data transformation (F) and the hyperparameter (H)  used during the actual attack within the victim phase is different from that used within the search process with Gradient \cite{gao2023not}, OPS \cite{guo2023temporal}, and FUS \cite{xia2022data} on Tiny-ImageNet datset. The horizontal coordinates are labeled as "0", "1", "2", "3", and "4", representing different combinations of data transforms and hyperparameters. All results are computed as the mean of five different runs.}
    \label{motivation_all}
\end{figure*}

\noindent\textbf{Effect of Different Hyperparameters (H) on Proxy Attack-based Sample Selection Efficiency.}
In this section, We thoroughly investigated the impact of different H on the effective data efficiency searched by FUS \cite{xia2022data}. Adhering to the aforementioned experimental framework, we configured the hyperparameter combination for the proxy poisoning attack during the sample selection phase as "V-16, SGD, 0.01". Subsequently, during the actual poisoning attack within the victim phase, we introduced variations by employing a range of H, including "V-13, SGD, 0.01", "V-13, Adam, 0.001", "V-16, Adam, 0.001", "R-18, SGD, 0.01", "R-18, Adam, 0.001", and "P-18, SGD, 0.01"\footnote{In this context, V-13, V-16, R-18, and P-18 denote VGG13, VGG16, ResNet18, and PreActResNet18, respectively. \label{foot1}}. Figure \ref{motivation_hyper} indicates that different H decrease the effectiveness of proxy attack-based sample selection methods. Once again, we observed anomalous occurrences when the parameters for both proxy searching and actual attacks closely converged, as illustrated by the scenario involving "V-16-SGD-0.01" proxy searching and subsequent "V-13-SGD-0.01" attacks within the victim phase.

\noindent\textbf{Summary}
Our findings underscore the susceptibility of proxy attack-based sample selection techniques to shifts in data transforms and hyperparameters deployed during the victim phase. Notably, SELF-FUS consistently outperforms FUS across various scenarios, with the exception of outliers observed in both Figure \ref{motivation_transform} and Figure \ref{motivation_hyper}. These outliers, marked by the "1+2" searching and "1" attacking in Figure \ref{motivation_transform}, as well as the "V-16-SGD-0.01" searching and "V-13-SGD-0.01" attacking in Figure \ref{motivation_hyper}, share closely aligned proxy and actual attack settings. To further validate our hypothesis, we continue to simulate more realistic attack scenarios that introduce disparities in both data transformations (F) and hyperparameters (H) between proxy and actual attacks.  Figure \ref{motivation_all} presents the ASR comparison between FUS and SELF-FUS, Gradient and SELF-Gradient, and OPS and SELF-OPS on the Tiny-ImageNet dataset. The notable differences in ASR highlight the significant impact of proxy attack-based methods on sample efficiency.

\begin{table*}
	\centering
	\scriptsize
 \vspace{-1.0em}
 \caption{The attack success rate under different optimizers on the CIFAR-10 dataset. All results are averaged over 5 different runs. The poisoning rates of experiments with trigger BadNets, Blended, ISSBA, and Optimized are $1\%$, $1\%$, $1\%$, and $0.5\%$, respectively. Among four select strategies, the best result is denoted in \textbf{boldface} while the \underline{underline} indicates the second-best result. The settings corresponding to the gray background are represented as the proxy poisoning attack within the attacker phase and the actual poisoning process within the victim phase being consistent in FUS.
	\label{tab:cifar10_optimizer}}
	\scalebox{1}{
		\begin{tabular}{cccccc|cccc|cccc}
			\toprule
			\multirow{3}{*}{\makecell*[c]{Trigger}} &\multirow{3}{*}{Optimizer} &\multicolumn{12}{c}{\makecell*[c]{Model}} \\ 
			\cline{3-14} 
			\specialrule{0em}{1pt}{1pt}
			&&
			\multicolumn{4}{c}{\makecell*[c]{VGG16}}&\multicolumn{4}{c}{\makecell*[c]{ResNet18}}&\multicolumn{4}{c}{\makecell*[c]{PreAct-ResNet18}}\\
			
			&&Random&PFS&FUS&FUS+PFS&Random&PFS&FUS &FUS+PFS&Random&PFS&FUS&FUS+PFS\\
			\midrule
			\multirow{5}{*}{BadNets}&SGD-0.01&0.916&\underline{0.966}&\cellcolor{mygray}0.956&\cellcolor{mygray}\textbf{0.972}&0.933&\underline{0.980}&0.970&\textbf{0.983}&0.931&\underline{0.980}&0.970&\textbf{0.983}\\
   &SGD-0.02&0.918&\underline{0.971}&0.962&\textbf{0.976}&0.938&\underline{0.983}&0.975&\textbf{0.986}&0.943&\underline{0.984}&0.976&\textbf{0.987}\\
   &Adam-0.0005&0.920&0.934&\underline{0.963}&\textbf{0.971}&0.945&0.960&\underline{0.975}&\textbf{0.985}&0.944&0.960&\underline{0.976}&\textbf{0.984}\\
   &Adam-0.001&0.928&\textbf{0.971}&\underline{0.967}&0.957&0.950&\underline{0.985}&0.977&\textbf{0.986}&0.954&\underline{0.982}&0.977&\textbf{0.985}\\
			\cline{2-14}
			\specialrule{0em}{1pt}{1pt}
			&Avg&0.921&0.961&\underline{0.962}&\textbf{0.969}&0.942&\underline{0.977}&0.974&\textbf{0.985}&0.943&\underline{0.977}&0.975&\textbf{0.985}\\
			\cmidrule{1-14} 
			\multirow{5}{*}{Blended}&SGD-0.01&0.827&0.883&\cellcolor{mygray}\underline{0.911}&\cellcolor{mygray}\textbf{0.933}&0.855&0.931&\underline{0.932}&\textbf{0.958}&0.864&\underline{0.926}&0.924&\textbf{0.956}\\
	&SGD-0.02&0.836&0.887&\underline{0.934}&\textbf{0.935}&0.881&0.939&\underline{0.950}&\textbf{0.968}&0.881&0.947&\underline{0.948}&\textbf{0.967}\\
   &Adam-0.0005&0.873&0.890&\textbf{0.924}&\underline{0.908}&0.897&0.952&\underline{0.953}&\textbf{0.974}&0.893&\underline{0.952}&0.940&\textbf{0.966}\\
   &Adam-0.001&0.886&\underline{0.910}&\textbf{0.937}&0.909&0.900&\underline{0.953}&0.951&\textbf{0.970}&0.896&\underline{0.949}&0.943&\textbf{0.962}\\
			\cline{2-14}
			\specialrule{0em}{1pt}{1pt}
			&Avg&0.856&0.891&\textbf{0.927}&\underline{0.921}&0.883&0.944&\underline{0.947}&\textbf{0.968}&0.884&\underline{0.944}&0.939&\textbf{0.963}\\
			\cmidrule{1-14} 
			\multirow{5}{*}{ISSBA}&SGD-0.01&0.741&0.801&\cellcolor{mygray}\underline{0.828}&\cellcolor{mygray}\textbf{0.856}&0.769&\underline{0.862}&{0.847}&\textbf{0.882}&0.779&\underline{0.851}&0.838&\textbf{0.868}\\
	&SGD-0.02&0.763&\underline{0.839}&{0.834}&\textbf{0.857}&0.806&\underline{0.874}&{0.870}&\textbf{0.889}&0.803&\underline{0.880}&{0.878}&\textbf{0.894}\\
   &Adam-0.0005&0.795&\underline{0.826}&{0.819}&\textbf{0.838}&0.822&\underline{0.891}&{0.885}&\textbf{0.902}&0.820&\underline{0.886}&0.872&\textbf{0.899}\\
   &Adam-0.001&0.811&\underline{0.862}&{0.858}&\textbf{0.869}&0.827&\underline{0.883}&0.875&\textbf{0.901}&0.822&\underline{0.879}&0.861&\textbf{0.902}\\
			\cline{2-14}
			\specialrule{0em}{1pt}{1pt}
			&Avg&0.778&0.832&\underline{0.835}&\textbf{0.855}&0.806&\underline{0.878}&{0.869}&\textbf{0.894}&0.806&\underline{0.874}&0.862&\textbf{0.891}\\
			\cmidrule{1-14} 
			\multirow{5}{*}{Optimized}&SGD-0.01&0.979&\underline{0.999}&\cellcolor{mygray}0.996&\cellcolor{mygray}\textbf{1.000}&0.979&\underline{0.999}&0.997&\textbf{1.000}&0.982&\underline{0.999}&0.997&\textbf{1.000}\\
			&SGD-0.02&0.979&\underline{0.999}&0.998&\textbf{1.000}&0.982&\underline{0.999}&0.997&\textbf{1.000}&0.984&\underline{0.999}&0.997&\textbf{1.000}\\
   &Adam-0.0005&0.975&\underline{0.997}&0.996&\textbf{1.000}&0.986&\underline{0.998}&0.996&\textbf{1.000}&0.982&\underline{0.998}&0.996&\textbf{1.000}\\
   &Adam-0.001&0.980&\underline{0.999}&0.997&\textbf{1.000}&0.985&\underline{0.999}&0.997&\textbf{1.000}&0.982&\underline{0.998}&0.994&\textbf{1.000}\\
			\cline{2-14}
			\specialrule{0em}{1pt}{1pt}
			&Avg&0.978&\underline{0.999}&0.997&\textbf{1.000}&0.983&\underline{0.999}&0.997&\textbf{1.000}&0.983&\underline{0.997}&0.996&\textbf{1.000}\\
			\bottomrule
	\end{tabular}}
\end{table*}

\begin{figure*}[t!]
    \vspace{-1em}
    \centering
    \includegraphics[width=\textwidth]{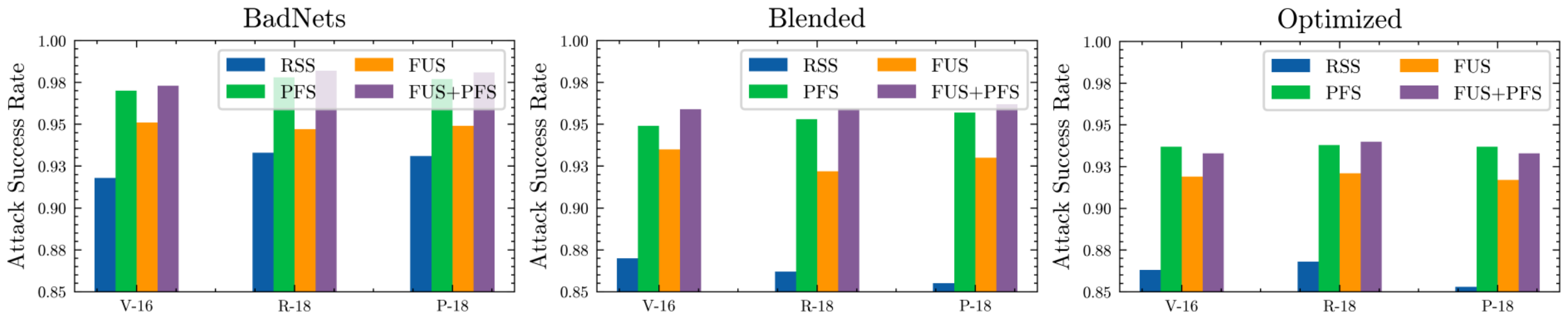}
    \caption{Average ASR of different data transforms and optimizers on the Tiny-ImageNet dataset.  Notably, PFS consistently outperforms FUS across all settings.}
    \label{imagenet_histogram}
\end{figure*}
\begin{table*}
	\centering
	\scriptsize
 \vspace{-1em}
 \caption{The attack success rate under different data transforms on the CIFAR-10 dataset. All results are averaged over 5 different runs. The poisoning rates of experiments with trigger BadNets, Blended, ISSBA, and Optimized are $1\%$, $1\%$, $1\%$, and $0.5\%$, respectively. Among four select strategies, the best result is denoted in \textbf{boldface} while the \underline{underline} indicates the second-best result. The settings corresponding to the gray background are represented as the proxy poisoning attack within the attacker phase and the actual poisoning process within the victim phase being consistent in FUS. 
	\label{tab:cifar10}}
	\scalebox{0.9}{
		\begin{tabular}{cccccc|cccc|cccc}
			\toprule
			\multirow{3}{*}{\makecell*[c]{Trigger}} &\multirow{3}{*}{Data Transform} &\multicolumn{12}{c}{\makecell*[c]{Model}} \\ 
			\cline{3-14} 
			\specialrule{0em}{1pt}{1pt}
			&&
			\multicolumn{4}{c}{\makecell*[c]{VGG16}}&\multicolumn{4}{c}{\makecell*[c]{ResNet18}}&\multicolumn{4}{c}{\makecell*[c]{PreAct-ResNet18}}\\
			
			&&Random&PFS&FUS&FUS+PFS&Random&PFS&FUS &FUS+PFS&Random&PFS&FUS&FUS+PFS\\
			\midrule
			\multirow{7}{*}{BadNets}&None&0.950&\underline{0.984}&0.978&\textbf{0.988}&0.964&\underline{0.992}&0.988&\textbf{0.994}&0.961&\underline{0.991}&0.985&\textbf{0.994}\\
			&RandomCrop&0.906&\underline{0.964}&0.954&\textbf{0.972}&0.918&\textbf{0.976}&0.966&\textbf{0.976}&0.927&\underline{0.977}&0.968&\textbf{0.981}\\
			&RandomHorizontalFlip&0.932&\underline{0.981}&0.971&\textbf{0.986}&0.964&\underline{0.989}&0.983&\textbf{0.992}&0.954&\underline{0.987}&0.981&\textbf{0.990}\\
			&RandomRotation&0.879&\underline{0.952}&0.939&\textbf{0.953}&0.909&\underline{0.972}&0.955&\textbf{0.976}&0.918&\underline{0.974}&0.961&\textbf{0.977}\\
			&ColorJitter&0.952&\underline{0.982}&0.975&\textbf{0.987}&0.967&\underline{0.993}&0.988&\textbf{0.995}&0.964&\underline{0.992}&0.985&\textbf{0.994}\\
			&{\makecell[c]{RandomCrop+ \\ RandomHorizontalFlip}}&0.916&\underline{0.966}&\cellcolor{mygray}0.956&\cellcolor{mygray}\textbf{0.972}&0.933&\underline{0.980}&0.970&\textbf{0.983}&0.931&\underline{0.980}&0.970&\textbf{0.983}\\
			\cline{2-14}
			\specialrule{0em}{1pt}{1pt}
			&Avg&0.923&\underline{0.972}&0.962&\textbf{0.976}&0.943&\underline{0.984}&0.975&\textbf{0.986}&0.943&\underline{0.984}&0.975&\textbf{0.987}\\
			\cmidrule{1-14} 
			\multirow{7}{*}{Blended}&None&0.795&0.856&\underline{0.861}&\textbf{0.889}&0.814&\underline{0.923}&0.885&\textbf{0.930}&0.867&\underline{0.949}&0.919&\textbf{0.957}\\
			&RandomCrop&0.792&0.830&\underline{0.865}&\textbf{0.874}&0.818&\underline{0.888}&0.885&\textbf{0.915}&0.821&\underline{0.903}&0.896&\textbf{0.921}\\
			&RandomHorizontalFlip&0.827&0.899&\underline{0.916}&\textbf{0.930}&0.867&\underline{0.948}&0.925&\textbf{0.959}&0.864&\underline{0.954}&0.923&\textbf{0.962}\\
			&RandomRotation&0.791&0.821&\underline{0.855}&\textbf{0.864}&0.833&\underline{0.897}&0.890&\textbf{0.926}&0.838&\underline{0.918}&0.902&\textbf{0.933}\\
			&ColorJitter&0.758&\underline{0.846}&0.826&\textbf{0.861}&0.758&\underline{0.846}&0.826&\textbf{0.921}&0.758&\underline{0.846}&0.826&\textbf{0.950}\\
			&{\makecell[c]{RandomCrop+ \\ RandomHorizontalFlip}}&0.827&0.883&\cellcolor{mygray}\underline{0.911}&\cellcolor{mygray}\textbf{0.933}&0.855&0.931&\underline{0.932}&\textbf{0.958}&0.864&\underline{0.926}&0.924&\textbf{0.956}\\
			\cline{2-14}
			\specialrule{0em}{1pt}{1pt}
			&Avg&0.798&0.856&\underline{0.872}&\textbf{0.892}&0.824&\underline{0.906}&0.891&\textbf{0.935}&0.835&\underline{0.916}&0.898&\textbf{0.947}\\
			\cmidrule{1-14} 
			
			\multirow{7}{*}{ISSBA}&None&0.712&\underline{0.782}&{0.779}&\textbf{0.795}&0.729&\underline{0.841}&0.801&\textbf{0.852}&0.779&\underline{0.862}&0.833&\textbf{0.882}\\
			&RandomCrop&0.708&\underline{0.779}&{0.776}&\textbf{0.802}&0.733&\underline{0.811}&0.809&\textbf{0.832}&0.744&\underline{0.824}&0.819&\textbf{0.852}\\
			&RandomHorizontalFlip&0.732&0.834&\underline{0.841}&\textbf{0.860}&0.792&\underline{0.883}&0.852&\textbf{0.885}&0.791&\underline{0.881}&0.849&\textbf{0.899}\\
			&RandomRotation&0.709&0.766&\underline{0.772}&\textbf{0.791}&0.753&\underline{0.819}&0.812&\textbf{0.847}&0.754&\underline{0.846}&0.828&\textbf{0.852}\\
			&ColorJitter&0.681&\underline{0.763}&0.731&\textbf{0.792}&0.689&\underline{0.772}&0.739&\textbf{0.838}&0.681&\underline{0.769}&0.750&\textbf{0.877}\\
			&{\makecell[c]{RandomCrop+ \\ RandomHorizontalFlip}}&0.750&0.802&\cellcolor{mygray}\underline{0.834}&\cellcolor{mygray}\textbf{0.858}&0.781&\underline{0.857}&{0.844}&\textbf{0.882}&0.783&\underline{0.842}&0.839&\textbf{0.869}\\
			\cline{2-14}
			\specialrule{0em}{1pt}{1pt}
			&Avg&0.715&0.788&\underline{0.789}&\textbf{0.816}&0.746&\underline{0.831}&0.810&\textbf{0.856}&0.755&\underline{0.837}&0.820&\textbf{0.872}\\
   \cmidrule{1-14} 
			
			\multirow{7}{*}{Optimized}&None&0.844&\underline{0.973}&0.910&\textbf{0.988}&0.840&\underline{0.952}&0.897&\textbf{0.968}&0.902&\underline{0.982}&0.940&\textbf{0.993}\\
			&RandomCrop&0.952&\underline{0.998}&0.990&\textbf{1.000}&0.966&\underline{0.999}&0.994&\textbf{1.000}&0.970&\underline{0.999}&0.993&\textbf{1.000}\\
			&RandomHorizontalFlip&0.922&\underline{0.990}&0.975&\textbf{0.998}&0.885&\underline{0.977}&0.945&\textbf{0.993}&0.932&\underline{0.994}&0.981&\textbf{0.998}\\
			&RandomRotation&0.844&\textbf{0.893}&0.873&\underline{0.874}&0.828&\textbf{0.930}&0.879&\underline{0.914}&0.863&\textbf{0.958}&0.918&\underline{0.956}\\
			&ColorJitter&0.875&\underline{0.962}&0.933&\textbf{0.991}&0.861&\underline{0.940}&0.918&\textbf{0.977}&0.911&\underline{0.985}&0.969&\textbf{0.995}\\
			&{\makecell[c]{RandomCrop+ \\ RandomHorizontalFlip}}&0.979&\underline{0.999}&\cellcolor{mygray}0.996&\cellcolor{mygray}\textbf{1.000}&0.979&\underline{0.999}&0.997&\textbf{1.000}&0.982&\underline{0.999}&0.997&\textbf{1.000}\\
			\cline{2-14}
			\specialrule{0em}{1pt}{1pt}
			&Avg&0.903&\underline{0.969}&0.946&\textbf{0.975}&0.893&\underline{0.966}&0.938&\textbf{0.975}&0.927&\underline{0.986}&0.966&\textbf{0.990}\\
			\bottomrule
	\end{tabular}}
\end{table*}
\section{Experiments}
We evaluate the effectiveness of PFS on datasets: CIFAR -10 (Sec. \ref{exp_cifar10}), Tiny-ImageNet (Sec. \ref{exp_tiny}), and CIFAR-100 (Sec. F of the \textbf{Supplementary Materials}). We also provide experimental settings in Sec. \ref{exp_setting}, comparison with recent sample selection methods in Sec. \ref{sec:state-of-the-art}, time consumption analysis in Sec. \ref{ana_tim_comsumption}, ablation study in Sec. \ref{exp_ablation}, and experiment against defense method in Sec. \ref{exp_deference}.

\subsection{Experimental Settings}
\label{exp_setting}
We conduct comprehensive experiments encompassing four triggers\footnote{Additionally, we consider the recent backdoor attacks such as WaNet \cite{nguyen2021wanet}, ISSBA \cite{li2021invisible}, and BATT \cite{xu2023batt} in Sec. \ref{sec:state-of-the-art}. } (BadNets \cite{gu2019badnets}, Blended \cite{chen2017targeted}, Optimized \cite{zhong2020backdoor}, and ISSBA \cite{li2021invisible}), six data transformations (None, RandomCrop, RandomHorizontalFlip, RandomRotation, ColorJitter, RandomCrop+RandomHorizontalFlip), and three different DNN architectures (VGG16 \cite{simonyan2014very}, ResNet18 \cite{he2016deep}, PreAct-ResNet18 \cite{he2016identity}). Furthermore, we explore the effects of four optimizers: SGD-0.01\footnote{SGD-0.01 implies the adoption of the SGD optimizer with an initial learning rate of 0.01.}, SGD-0.02, Adam-0.001, and Adam-0.002. The employed feature extractor $E(\cdot)$ is ResNet18, pre-trained on the specific dataset using the SGD optimizer with an initial learning rate of 0.01, momentum of 0.9, and weight decay of 5e-4, along with two standard data augmentations: random crop and random horizontal flip. Each training iteration spans 70 epochs, conducted with a batch size of 256. In addition, Sec. \ref{sec:exp_pretrained} presents experiments involving various pre-trained extractors on the CIFAR-10 dataset. Our findings reveal that even utilizing a commonly used ImageNet pre-trained extractor yields notable enhancements compared to the baseline.


We perform a comprehensive comparative analysis, evaluating our method against the widely employed random selection (Random) \cite{gu2019badnets,chen2017targeted} and FUS \cite{xia2022data} approaches. In the FUS search procedure, we utilize the VGG16 architecture, coupled with RandomCrop+RandomHorizontalFlip data transformations and SGD-0.01 as the optimizer to establish the proxy attack. Conversely, for our proposed method, we utilize pre-trained ResNet models for similarity measurement and set the diversity hyperparameter $m$ to 10 across all three datasets. All other hyperparameters remain consistent with those outlined in Sec. \ref{sec:exp_setting}. Additionally, Sec. \ref{sec:state-of-the-art} shows the comparison with recent sample selection methods (FUS \cite{xia2022data}, OPS \cite{guo2023temporal}, Gradient\cite{gao2023not}, RD Score\cite{wu2023computation}) on the Tiny-ImageNet dataset.


\subsection{Experiments on CIFAR-10 Dataset}
\label{exp_cifar10}
\noindent\textbf{Results on different data transforms and architectures.}
Table \ref{tab:cifar10} provides a comprehensive overview of the ASR for BadNets, Blended, ISSBA, and Optimized triggers, considering various data transforms and architectures. Constructing the proxy attack for FUS, we employ VGG16 with RandomCrop+RandomHorizontalFlip and SGD-0.01 as the optimizer. Our proposed PFS method significantly enhances the poisoning efficacy in backdoor attacks. Across all 72 configurations, the ASR of poisoning samples selected through our approach consistently surpasses that of randomly selected samples at identical poisoning rates. On average, our method achieves a improvement of 0.044, 0.074, 0.08, and 0.066 in terms of ASR for BadNets, Blended, ISSBA, and Optimized triggers, respectively. In general, our method outperforms FUS, demonstrating greater ASR in 63 of the 72 scenarios. Notably, FUS searches necessitate significantly more time compared to our method, often by hundreds or even thousands of times. Furthermore, when the model is VGG16 and the trigger is set to Blended, FUS surpasses our PFS. This can be attributed to two factors: (i) Both the proxy attack model employed in FUS and the model utilized in the victim phase are VGG16; (ii) In comparison to BadNets and Optimized triggers, the Blended trigger increases the similarity distance between benign and corresponding poisoning samples, thereby marginally impacting the performance of our proposed PFS.
The combination of FUS and PFS, denoted as FUS+PFS, excels in 69 out of the 72 scenarios. This outcome stems from PFS effectively eliminating inefficient samples, consequently decreasing the search space for FUS.

\begin{table*}
	\centering
	\scriptsize
 \caption{The attack success rate on Tiny-ImageNet dataset. All results are averaged over 5 different runs. The poisoning rates of experiments with trigger BadNets, Blended, and Optimized are $1\%$, $1\%$, and $0.5\%$, respectively. Among four select strategies, the best result is denoted in \textbf{boldface} while the \underline{underline} indicates the second-best result. The settings corresponding to the gray background are represented as the proxy poisoning attack within the attacker phase and the actual poisoning process within the victim phase being consistent in FUS.
	\label{tab:imagenet}}
	\scalebox{1}{
		\begin{tabular}{cccccc|cccc|cccc}
			\toprule
			\multirow{3}{*}{\makecell*[c]{Trigger}} &\multirow{3}{*}{Data Transform} &\multicolumn{12}{c}{\makecell*[c]{Model}} \\ 
			\cline{3-14} 
			\specialrule{0em}{1pt}{1pt}
			&&
			\multicolumn{4}{c}{\makecell*[c]{VGG16}}&\multicolumn{4}{c}{\makecell*[c]{ResNet18}}&\multicolumn{4}{c}{\makecell*[c]{PreAct-ResNet18}}\\
			&&Random&PFS&FUS&FUS+PFS&Random&PFS&FUS &FUS+PFS&Random&PFS&FUS&FUS+PFS\\
			\midrule
			\multirow{7}{*}{BadNets}&None&0.952&\textbf{0.990}&0.983&\underline{0.989}&0.960&\textbf{0.992}&0.985&\underline{0.991}&0.956&\textbf{0.991}&0.981&\underline{0.989}\\
			&RandomCrop&0.893&\textbf{0.959}&0.950&\underline{0.958}&0.917&\underline{0.970}&0.963&\textbf{0.973}&0.904&\textbf{0.973}&0.958&\underline{0.971}\\
			&RandomHorizontalFlip&0.952&\textbf{0.989}&\underline{0.984}&\textbf{0.989}&0.963&\textbf{0.991}&0.981&\underline{0.989}&0.947&\underline{0.983}&0.974&\textbf{0.985}\\
			&RandomRotation&0.881&0.932&\underline{0.937}&\textbf{0.944}&0.912&0.959&\underline{0.964}&\textbf{0.966}&0.910&0.957&\underline{0.963}&\textbf{0.970}\\
			&ColorJitter&0.956&\textbf{0.990}&\underline{0.985}&\textbf{0.990}&0.966&\textbf{0.993}&0.986&\underline{0.992}&0.956&\textbf{0.990}&0.983&\underline{0.988}\\
			&{\makecell[c]{RandomCrop+ \\ RandomHorizontalFlip}}&0.900&0.943&\cellcolor{mygray}\textbf{0.955}&\cellcolor{mygray}\underline{0.950}&0.917&\textbf{0.977}&0.967&\underline{0.975}&0.924&\underline{0.974}&0.965&\textbf{0.978}\\
			\cline{2-14}
			\specialrule{0em}{1pt}{1pt}
			&Avg&0.922&\underline{0.967}&0.966&\textbf{0.970}&0.939&\underline{0.980}&0.974&\textbf{0.981}&0.933&\underline{0.978}&0.971&\textbf{0.980}\\
			
			\cmidrule{1-14} 
			\multirow{7}{*}{Blended}&None&0.889&\underline{0.964}&0.947&\textbf{0.975}&0.876&\underline{0.965}&0.930&\textbf{0.973}&0.873&\underline{0.969}&0.941&\textbf{0.978}\\
			&RandomCrop&0.836&\underline{0.928}&0.920&\textbf{0.942}&0.835&\underline{0.931}&0.914&\textbf{0.936}&0.822&\underline{0.927}&0.906&\textbf{0.932}\\
			&RandomHorizontalFlip&0.905&\underline{0.966}&0.964&\textbf{0.980}&0.888&\underline{0.964}&0.950&\textbf{0.977}&0.886&\underline{0.965}&0.954&\textbf{0.981}\\
			&RandomRotation&0.865&0.931&\underline{0.946}&\textbf{0.953}&0.872&\underline{0.945}&0.942&\textbf{0.966}&0.870&\underline{0.950}&0.941&\textbf{0.966}\\
			&ColorJitter&0.883&\underline{0.958}&0.945&\textbf{0.964}&0.871&\underline{0.956}&0.917&\textbf{0.962}&0.872&\underline{0.967}&0.931&\textbf{0.971}\\
			&{\makecell[c]{RandomCrop+ \\ RandomHorizontalFlip}}&0.850&0.931&\cellcolor{mygray}\underline{0.940}&\cellcolor{mygray}\textbf{0.958}&0.840&\underline{0.929}&0.919&\textbf{0.942}&0.828&\underline{0.931}&0.912&\textbf{0.940}\\
			\cline{2-14}
			\specialrule{0em}{1pt}{1pt}
			&Avg&0.871&\underline{0.946}&0.944&\textbf{0.962}&0.864&\underline{0.948}&0.929&\textbf{0.959}&0.859&\underline{0.952}&0.931&\textbf{0.961}\\
			\cmidrule{1-14} 
			\multirow{7}{*}{Optimized}&None&0.907&\underline{0.981}&0.956&\textbf{0.984}&0.875&\textbf{0.968}&\underline{0.934}&\textbf{0.968}&0.862&\textbf{0.970}&0.935&\underline{0.967}\\
			&RandomCrop&0.897&\underline{0.976}&0.959&\textbf{0.980}&0.886&\underline{0.968}&0.950&\textbf{0.972}&0.869&\textbf{0.958}&0.945&\underline{0.951}\\
			&RandomHorizontalFlip&0.903&\underline{0.979}&0.956&\textbf{0.985}&0.892&\textbf{0.976}&\underline{0.950}&\textbf{0.976}&0.877&\textbf{0.975}&0.948&\underline{0.974}\\
			&RandomRotation&0.899&\textbf{0.977}&0.957&\underline{0.972}&0.897&\textbf{0.978}&0.956&\underline{0.972}&0.888&\textbf{0.973}&0.948&\underline{0.956}\\
			&ColorJitter&0.695&\underline{0.714}&\textbf{0.744}&0.707&0.769&0.773&\textbf{0.803}&\underline{0.775}&0.772&0.779&\textbf{0.816}&\underline{0.781}\\
			&{\makecell[c]{RandomCrop+ \\ RandomHorizontalFlip}}&0.892&\underline{0.973}&\cellcolor{mygray}0.960&\cellcolor{mygray}\textbf{0.974}&0.890&\underline{0.964}&0.952&\textbf{0.968}&0.870&\textbf{0.963}&0.949&\underline{0.957}\\
			\cline{2-14}
			\specialrule{0em}{1pt}{1pt}
			&Avg&0.866&\underline{0.933}&0.922&\textbf{0.934}&0.868&\underline{0.938}&0.924&\textbf{0.939}&0.856&\textbf{0.936}&0.924&\underline{0.931}\\
			\bottomrule
	\end{tabular}}
\end{table*}

\begin{table}
	\caption{The comparison with recent sample selection methods on the Tiny-ImageNet dataset. All results are computed the mean by different data transforms. The poisoning rate of different poisoned attacks is set to $1\%$.
	\label{tab:sota}}
	\centering
	\scalebox{1.1}{
		\begin{tabular}{ccccc}
			\toprule
			\multirow{2}{*}{Sample Selection} &\multicolumn{4}{c}{\makecell*[c]{Attacks}} \\ 
			\specialrule{0em}{1pt}{1pt}
			&
			Blended&WaNet&ISSBA&BATT\\
			\midrule
      \specialrule{0em}{1pt}{1pt}
			  \cellcolor[gray]{0.85} RSS&\cellcolor[gray]{0.85}0.864&\cellcolor[gray]{0.85}0.813&\cellcolor[gray]{0.85}0.764&\cellcolor[gray]{0.85}0.852\\
            FUS \cite{xia2022data}&0.929&0.877&0.838&0.901\\
 \cellcolor[gray]{0.85} OPS \cite{guo2023temporal} &\cellcolor[gray]{0.85}0.882&\cellcolor[gray]{0.85}0.843&\cellcolor[gray]{0.85}0.792&\cellcolor[gray]{0.85}0.869\\
 Gradient\cite{gao2023not}&0.901&0.851&0.808&0.880\\
 \cellcolor[gray]{0.85}RD Score\cite{wu2023computation}&0.913\cellcolor[gray]{0.85}&\cellcolor[gray]{0.85}0.847&\cellcolor[gray]{0.85}0.806&\cellcolor[gray]{0.85}0.893\\
            PFS (Ours)&\textbf{0.948}&\textbf{0.893}&\textbf{0.861}&\textbf{0.935}\\

			\bottomrule
	\end{tabular}}
\end{table}
\noindent\textbf{Results on different optimizer and architectures.}
Table \ref{tab:cifar10_optimizer} presents the ASR for the BadNets, Blended, ISSBA, and Optimized triggers, considering different optimizers and architectures. We utilize VGG16 with RandomCrop+RandomHorizontalFlip data transformations and SGD-0.01 as the optimizer for constructing the proxy attack in FUS. The outcomes underscore the effectiveness of PFS method. Firstly, across all 48 configurations, PFS consistently achieves higher ASR than random selection, with average improvements of  0.0437, 0.052, 0.065, and 0.017 for the BadNets, Blended, ISSBA and Optimized triggers, respectively. Secondly, PFS outperforms FUS in 36 out of the 48 cases. Similar to previous experiments, when the model is VGG16 and the trigger is set to Blended, FUS outperforms PFS. Notably, the combination of our method with FUS, referred to as FUS+PFS, achieves the best performance in 45 out of the 48 settings.


The \textbf{Supplementary Materials} contain additional experiments aimed at further evaluating the effectiveness and robustness of our proposed strategy. Specifically, in Sec. C and Sec. D, we provide the ASR and benign accuracy (BA) under poisoning rates of $1.5\%$ and $2\%$ on the CIFAR-10 dataset, respectively. The outcomes illustrate the robustness of our method across different poisoning rates while maintaining benign accuracy. Moreover, in Sec. E, we present the ASR for an additional target class (class 3) on the CIFAR-10 dataset. Overall, our results consistently affirm the effectiveness of the proposed pfs and support our hypothesis that the efficiency of backdoor attacks depends on the interaction between individual similarity and set diversity.



\subsection{Experiments on Tiny-ImageNet Dataset}
\label{exp_tiny}

\noindent\textbf{Results on different data transforms and architectures.}
Table~\ref{tab:imagenet} shows that PFS outperforms the random method in terms of ASR of poisoning samples on Tiny-ImageNet, with average boosts of 0.054, 0.094, and 0.072 on BadNets, Blended, and Optimized triggers, respectively. In comparison to FUS, PFS outperforms FUS in 45 out of 54 cases. In contrast to experiments on the CIFAR-10 dataset, our proposed PFS method consistently achieves higher average ASR than FUS across different data transformations on the Tiny-ImageNet dataset. When the two methods are combined, the best results are achieved in 34 out of 54 settings, indicating the superiority of our method. These findings provide strong evidence supporting the effectiveness of our proposed method on the scenarios involving extensive datasets.

\noindent\textbf{Results on different data transforms, optimizer, and architectures.}
  To further illustrate the effectiveness of our proposed PFS, Figure \ref{imagenet_histogram} visually shows the average ASR across various data transforms and optimizers on the Tiny-ImageNet dataset. The findings indicate that all advanced sample selection strategies (FUS, PFS, and FUS+PFS) achieve notably superior results compared to the random selection strategy (RSS). Furthermore, our proposed PFS consistently achieves better ASR compared to FUS across all settings.


\begin{figure*}[t!]
    \setlength{\belowcaptionskip}{-0.5cm}
	\centering
	\includegraphics[width=0.9\linewidth]{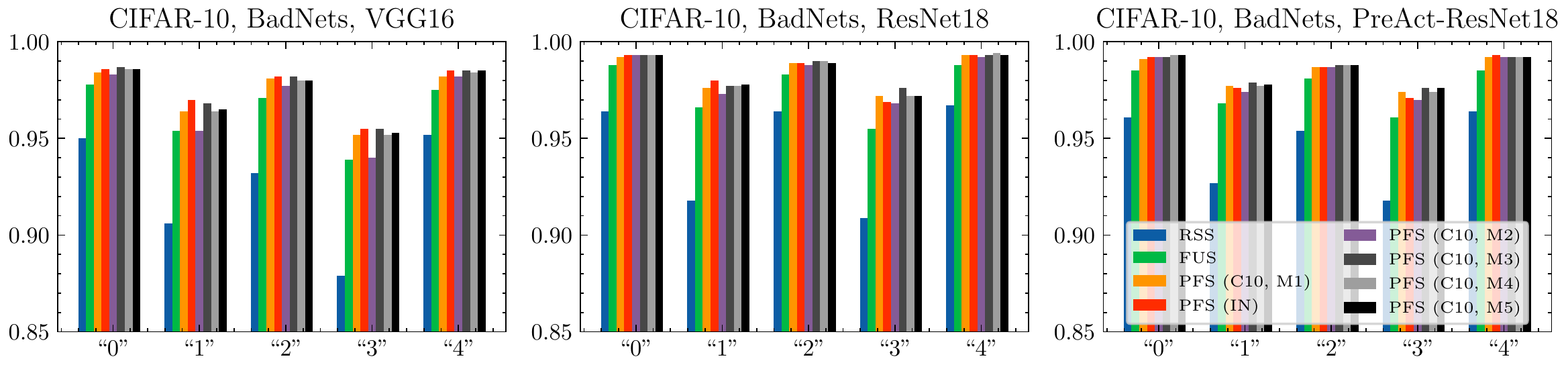}
	\caption{The ASR among different data transforms ("0", "1", "2", "3", and "4" are the same as that on Figure 2) on the CIFAR-10 dataset. we use different pre-trained feature extractor to construct our PFS. C10: CIFAR-10. IN: ImageNet. M1: R18-SGD-RandFlip+RandCrop.  M2: R18-Adam-RandRotation. M3: R18-SGD-RandRotation. M4: V16-SGD-RandRotation. M5: V16-SGD-None.}
	\label{fig:result_pre_trained}
\end{figure*}

\begin{table}
\centering
\scriptsize
\caption{Running time (sec) of different selection methods on the CIFAR-10 and Tiny-ImageNet dataset with an NVIDIA A100 GPU.
\label{tab:run_time}}
\scalebox{1}{
\begin{tabular}{ccc}
\toprule
Dataset&CIFAR-10&Tiny-ImageNet \\ 
\midrule
PFS (w/o pre-training) &17&19\\
PFS (w pre-training) &562&6784\\
FUS \cite{xia2022data}&9150&106284\\
Gradient \cite{gao2023not}&623&7395\\
OPS \cite{guo2023temporal}& 635&7487  \\
 
\bottomrule
\end{tabular}}
\end{table}
\subsection{Comparison with Recent Sample Selection Strategies}
\label{sec:state-of-the-art}
Table \ref{tab:sota} presents a comparative analysis between our PFS and recent sample selection strategies (FUS \cite{xia2022data}, OPS \cite{guo2023temporal}, Gradient\cite{gao2023not}, and RD Score\cite{wu2023computation}) on the Tiny-ImageNet dataset. Various recent triggers (Blended, WaNet, ISSBA, and BATT) are considered in this experiment. The results demonstrate that all sample selection strategies exhibit superior ASR compared to the random selection strategy (RSS), with our proposed PFS consistently demonstrating the highest sample efficiency across all settings when compared to other recent sample selection strategies.

\subsection{Experiments on Different Pre-trained Feature Extractor}
\label{sec:exp_pretrained}

Although both FUS and PFS rely on a proxy task, there are essential differences between proxy tasks. FUS relies on a proxy attack task closely resembling the actual attack. Nevertheless, due to the inherent shortcuts of backdoor learning, overfitting to the data transformation and hyperparameter of proxy attack significantly impact actual injection efficiency. In contrast, our PFS selects efficient samples considering individual similarity and set diversity. As a result, PFS operates independently of the specific attack deployed during the victim phase, with minimal impact from data transformation and hyperparameters within the pre-trained feature extractor on sample efficiency during this phase. This independence is evidenced in our experiments with various pre-trained extractors, as illustrated in Figure \ref{fig:result_pre_trained}. Remarkably, even a common extractor pre-trained on ImageNet results in considerable improvements compared to the baseline.

\subsection{Analysis of Time Consumption}
\label{ana_tim_comsumption}


Table~\ref{tab:run_time} presents a comparison of the running times of our PFS with those of FUS, Gradient, and OPS. PFS demonstrates remarkable speed, being $538\times$ faster than FUS on CIFAR-10 dataset. Even with pre-training time included, our method is still approximately $16\times$ faster on CIFAR-10 dataset. Importantly, our method utilizes the \textbf{same} pre-trained model for different triggers, while FUS needs to conduct a complete search from the beginning for each trigger. This advantage makes our method a practical choice for situations where efficiency is crucial.


\subsection{Ablation Study}
\label{exp_ablation}

\begin{figure}[t!]
    \setlength{\belowcaptionskip}{-0.5cm}
	\centering
	\includegraphics[width=0.9\linewidth]{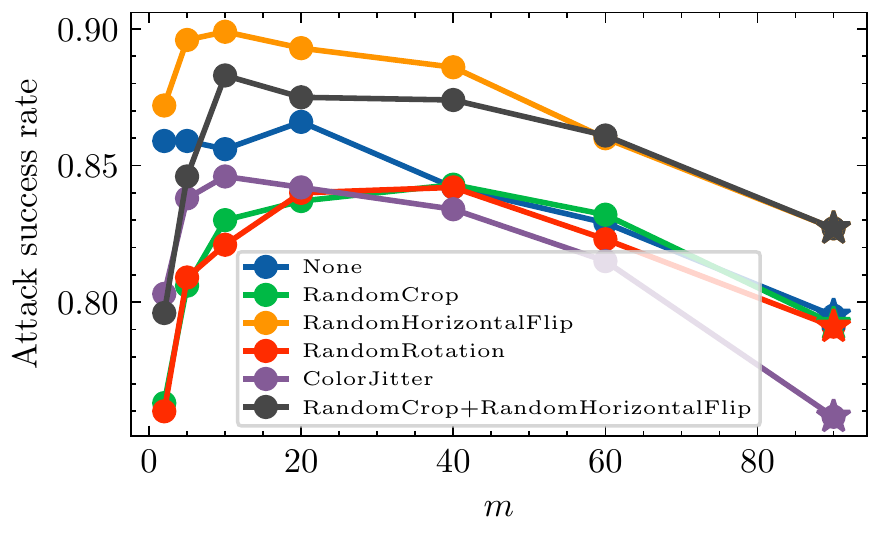}
	\caption{Ablation study of diversity rate $m$ on the CIFAR-10 dataset, where $\star$ indicate the results of Random Sampling.}
	\label{fig:ablation}
\end{figure}

We conduct an ablation study on the hyperparameter $m$, which determines the diversity of the poisoning sample set. The results are shown in \figurename~\ref{fig:ablation}, and the ASR demonstrate similar trends under different data transforms. Initially, as diversity gradually increases, the ASR also increase. However, too large of a value for $m$ can reduce the role of similarity and weaken the attack. These findings highlight the importance of balancing similarity and diversity for efficient poisoning samples, supporting our claim in Sec. \ref{sec:Forensic}. In our PFS, we set $m=10$ in all experimental settings.
 
\subsection{Experiment with Backdoor Defense.}
\label{exp_deference}

\begin{figure}[t!]
    \setlength{\belowcaptionskip}{-0.5cm}
	\centering
	\includegraphics[width=0.9\linewidth]{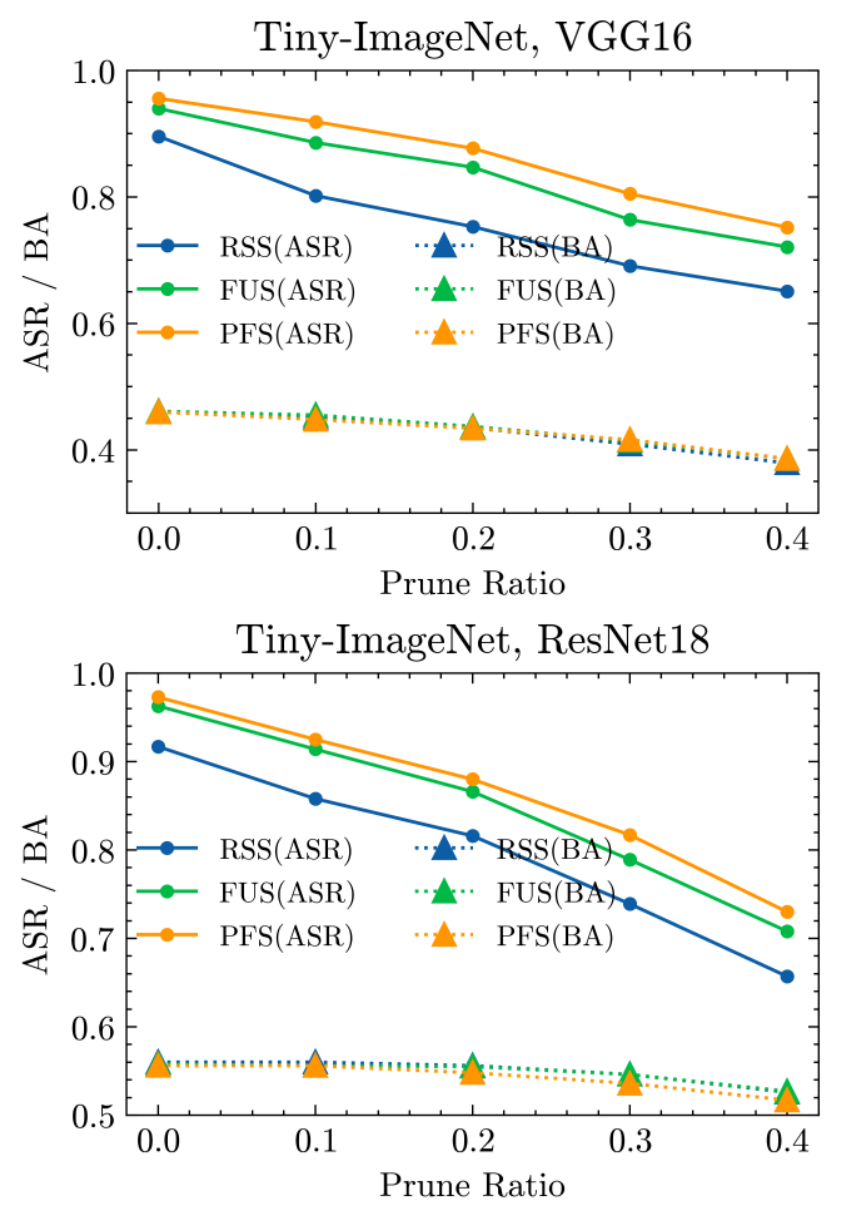}
	\caption{Defense results of pruning \cite{liu2018fine} on the Tiny-ImageNet dataset, where the number of clean samples owned by the defender is 500, and the poisoning ratio is $1\%$.}
	\label{fig:deference_pruning}
\end{figure}

\begin{figure}[t!]
    \setlength{\belowcaptionskip}{-0.5cm}
	\centering
	\includegraphics[width=0.9\linewidth]{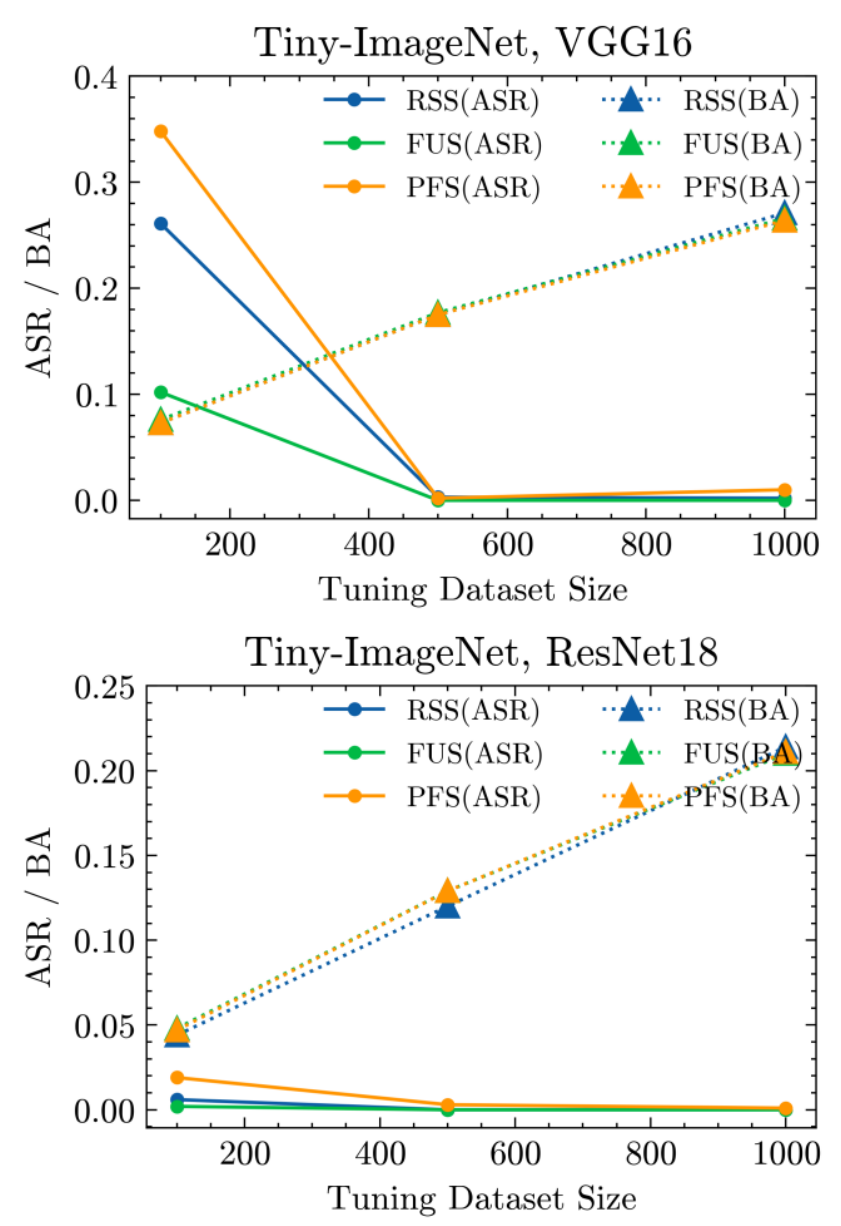}
	\caption{Defense results of FST \cite{min2023towards} on the Tiny-ImageNet dataset, where the number of clean samples owned by the defender are 100, 500, and 1000. The poisoning ratio is set to $1\%$. }
	\label{fig:deference_tuning}
\end{figure}

We evaluate the robustness of PFS against backdoor defense using the pruning-based defense method \cite{liu2018fine}, the tune-based defense method FST \cite{min2023towards}, and input-level backdoor detection SCALE-UP \cite{guo2023scale}. 
Pruning \cite{liu2018fine} is a widely used defense mechanism. It achieves this by pruning neurons that remain inactive in response to benign inputs, thereby neutralizing the effectiveness of the backdoor behavior. FST \cite{min2023towards} is a tuning-based backdoor purification method. It operates by inducing feature shifts through deliberate adjustments to the classifier weights, effectively diverging them from the initially compromised weights. SCALE-UP \cite{guo2023scale} only requires the predicted labels to realize the effective black-box input-level backdoor detection. Specifically, it identifies and filters malicious testing samples by analyzing their prediction consistency during the pixel-wise amplification process. Furthermore, defender adopts the filtered benign samples to train the model. 
As depicted in Figure \ref{fig:deference_pruning}, PFS demonstrates superior robustness against Pruning \cite{liu2018fine} compared to RSS and FUS. Similarly, Figure \ref{fig:deference_tuning} illustrates that our proposed PFS effectively against the tune-based defense method FST \cite{min2023towards} compared to RSS and FUS. It is noteworthy that while FST \cite{min2023towards} demonstrates superior defense capabilities compared to Pruning \cite{liu2018fine}, it comes at the expense of a substantial reduction in benign accuracy. Additionally, we adopt the data-free scaled prediction consistency analysis to evaluate the effectiveness of different attack methods against the SCALE-UP defense method \cite{guo2023scale}. As shown in Table \ref{tab:scale-up}, our proposed PFS method demonstrates superior performance in countering the backdoor detection method SCALE-UP compared to other selection baselines.




\begin{table}
\centering
\caption{Defense results (ASR) of SCALE-UP on the Tiny ImageNet dataset. The poisoning ratio is set to 1\%.
\label{tab:scale-up}}
\scalebox{1}{
\begin{tabular}{ccccc}
\toprule
Attack&BadNets&Blended&WaNet&ISSBA \\ 
\midrule
RSS&0.327&0.592&0.285&0.233\\
Gradient&0.363&0.633&0.348&0.294\\
FUS&0.411&0.661&0.384&0.316\\
PFS&0.452&0.685&0.415&0.337\\
\bottomrule
\end{tabular}}
\end{table}

\section{Conclusion and Limitations}

\noindent\textbf{Conclusion.} 
This paper contributes empirical insights into two interesting phenomena associated with data efficiency in backdoor attacks. Firstly, we highlight the performance degeneration experienced by sample selection methods relying on proxy attacks when discrepancies arise between proxy attacks within attacker phase and actual poisoning processes within the victim phase. Furthermore, we argue that the significance of both the similarity between benign and corresponding poisoning samples and the diversity within the poisoning sample set as essential factors for efficient data in backdoor attacks. Leveraging these observations, we introduce a simple yet efficient proxy attack-free sample selection strategy that is driven by the similarity between benign and corresponding poisoning samples. Our method significantly improves poisoning efficiency across benchmark datasets with minimal additional cost.

\noindent\textbf{Limitations and Future Work.} 
(i) While the proposed PFS has demonstrated efficient data selection in backdoor attacks, we note that similarity between benign and corresponding poisoned samples does not necessarily guarantee efficiency, as similarity only filters out non-effective samples. In the future, it would be desirable to identify an unified indicator that positively correlates with data efficiency in backdoor attacks. (ii) Moreover, our proposed PFS is customized specifically for dirty-label backdoor attacks. In future research, we intend to explore efficient sample selection strategies designed specifically for clean-label backdoor attacks. (iii) Our proposed PFS does not show significant improvement over the FUS method against backdoor defenses. Therefore, we will investigate more efficient sample selection methods to enhance efficiency against backdoor defenses in future studies.

\section*{Acknowledgments}
The work was supported by the National Natural Science Foundation of China under Grands U19B2044, U22B2062, 62172232, and 61836011; by the Jiangsu Basic Research Programs-Natural Science Foundation under grant numbers BK20200039; by the Collaborative Innovation Center of Atmospheric Environment and Equipment Technology fund.


{\small
\bibliographystyle{ieee_fullname}
\bibliography{egbib}
}
 \vspace{-10mm}
\begin{IEEEbiography}[{\includegraphics[width=1in,height=1.25in,clip,keepaspectratio]{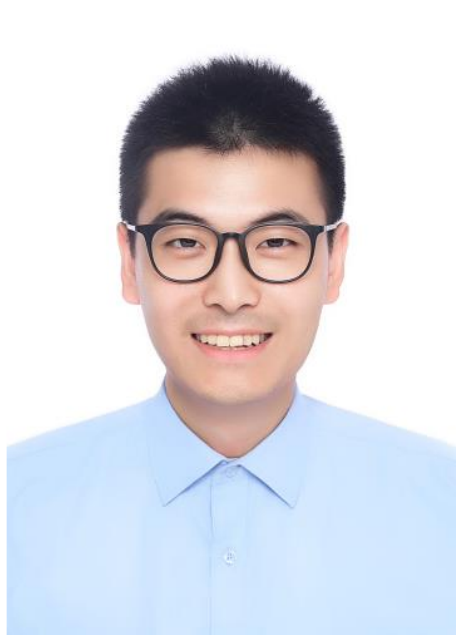}}]{Ziqiang Li}
		received the B.E. degree from University
		of Science and Technology of China (USTC), Hefei,
		China, in 2019, and the Ph.D degree from the University of Science and Technology
of China (USTC), Hefei, China, in 2024. He is
currently a Lecturer with the Engineering Research Center of Digital Forensics, Ministry of Education, Nanjing University of Information Science and Technology. His research interests include
		deep generative models, AI Security, and machine learning.
	\end{IEEEbiography}
 \vspace{-10mm}
\begin{IEEEbiography}[{\includegraphics[width=1in,height=1.25in,clip,keepaspectratio]{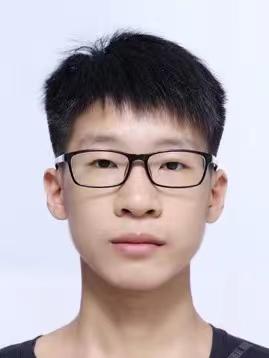}}]{Hong Sun}
		received the B.E. degree in electronic and information engineering from Dalian University of Technology (DUT), Dalian, China, in 2021, and the Master degree in Electronics and Communication Engineering from University of Science and Technology of China (USTC), Hefei, China, in 2024. His research interests include backdoor learning, generative adversarial
networks, and machine learning.
	\end{IEEEbiography}
 \vspace{-10mm}
\begin{IEEEbiography}[{\includegraphics[width=1in,height=1.25in,clip,keepaspectratio]{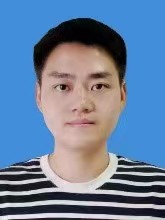}}]{Pengfei Xia}
		received the B.E. degree from
China University of Mining and Technology
(CUMT), Xuzhou, China, in 2015 and received
the Ph.D. degree from University of Science and
Technology of China (USTC), Hefei, China, in 2023. His
research interests include adversarial examples,
backdoor learning, and secure deep learning.
	\end{IEEEbiography}
  \vspace{-10mm}
  \begin{IEEEbiography}[{\includegraphics[width=1in,height=1.25in,clip,keepaspectratio]{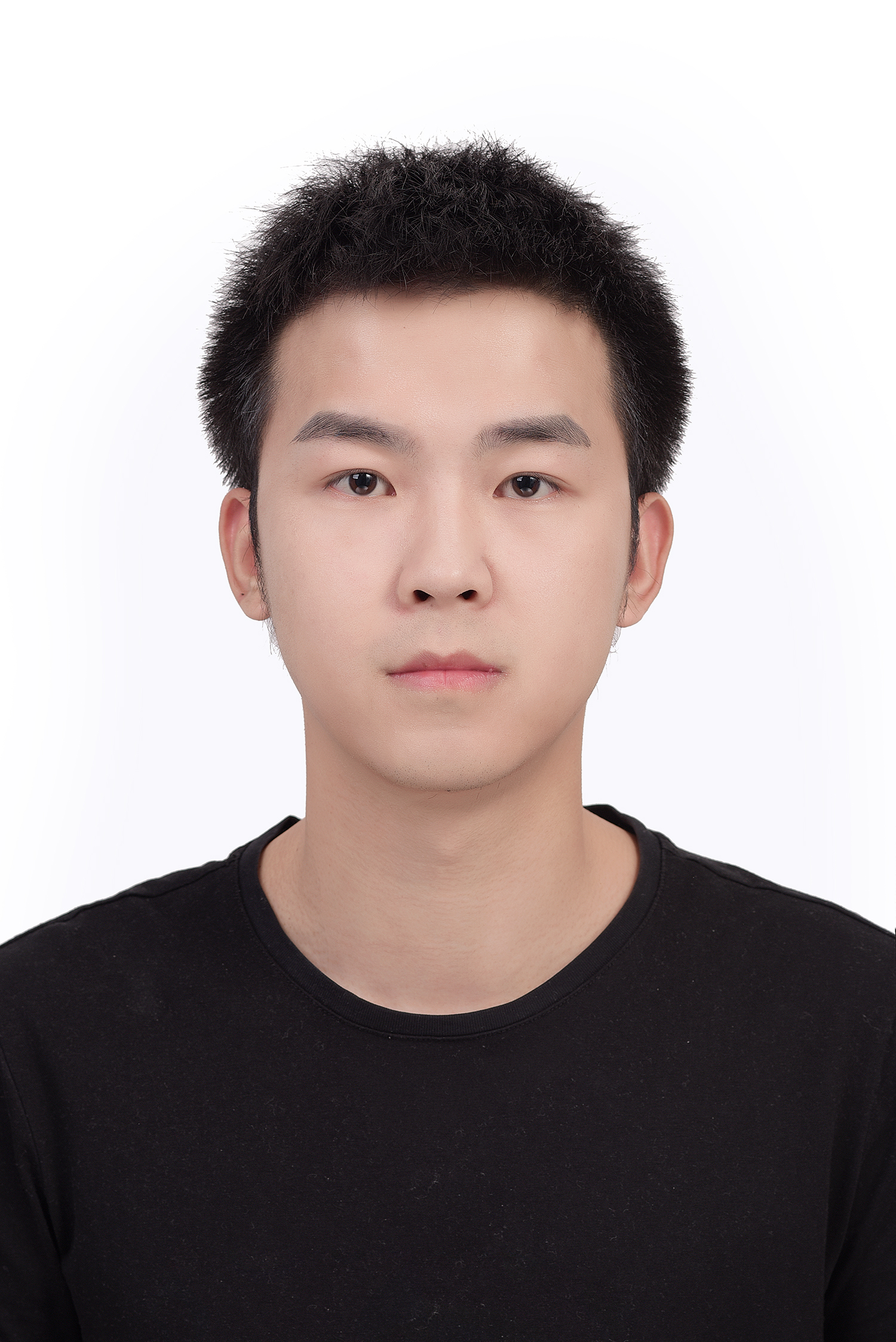}}]{Beihao Xia}
received the B.E. degree from the College of Computer Science and Electronic Engineering, Hunan University, Changsha, China, in 2015, and the Ph.D degree from the School of Electronic Information and Communications, Huazhong University of Sciences and Technology, Wuhan, China, in 2023. He is currently a Post-doctoral with Huazhong University of Sciences and Technology, Wuhan, China.  His current research interests include computer vision and machine learning.
\end{IEEEbiography}
 \vspace{-10mm}
 \begin{IEEEbiography}[{\includegraphics[width=1in,height=1.25in,clip,keepaspectratio]{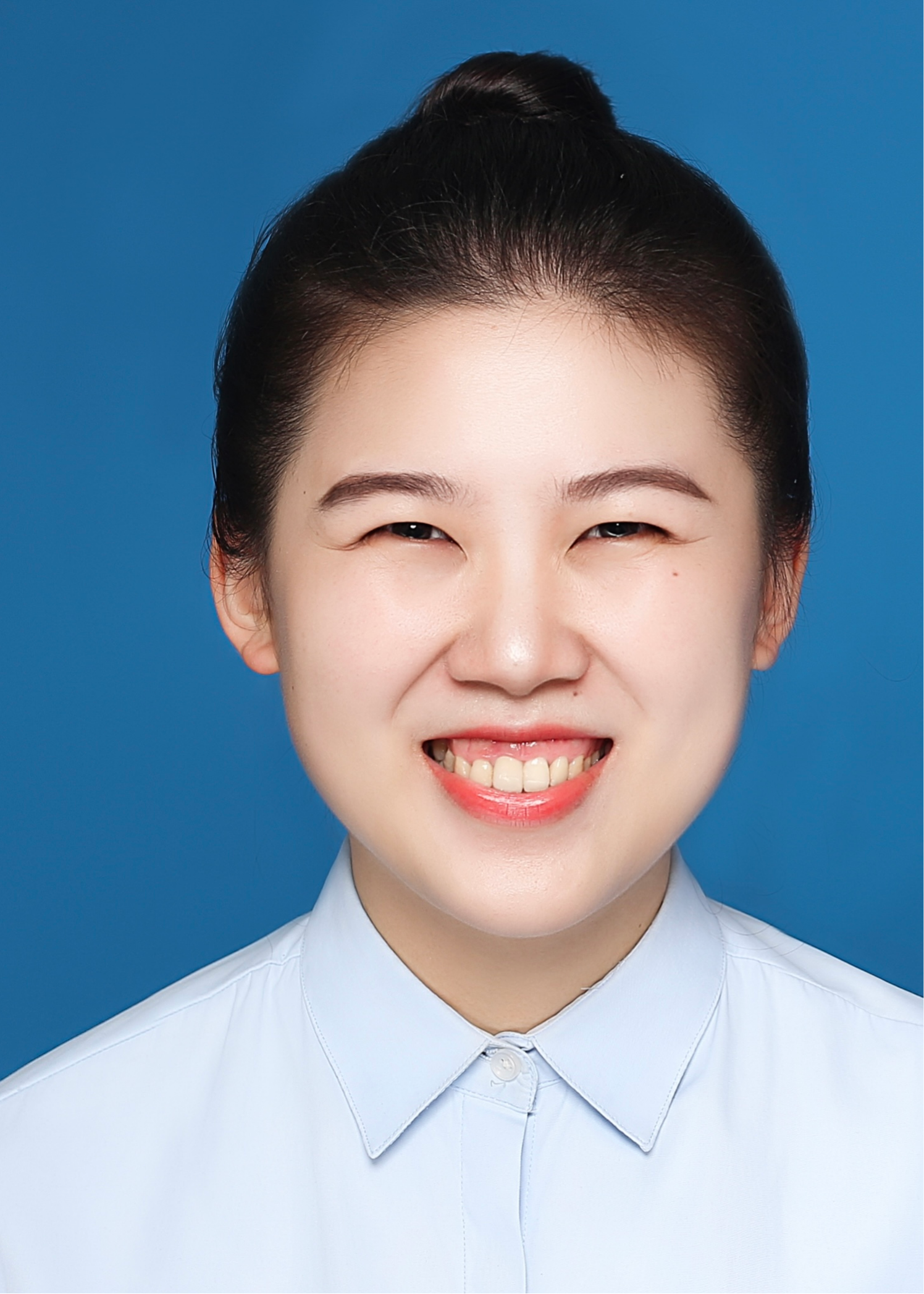}}]{Xue Rui}
received the B.E. degree from Chong Qing University (CQU), Chongqing,
		China, in 2018 and the  Ph.D degree from University of Science and Technology of China
		(USTC), Hefei, China, in 2024. She
		 is
currently a Lecturer with the Nanjing University of Information Science and Technology. Her research interests include
		computer vision,  remote sensing visual perception and earth vision.
\end{IEEEbiography}
 \vspace{-10mm}
 \begin{IEEEbiography}[{\includegraphics[width=1in,height=1.25in,clip,keepaspectratio]{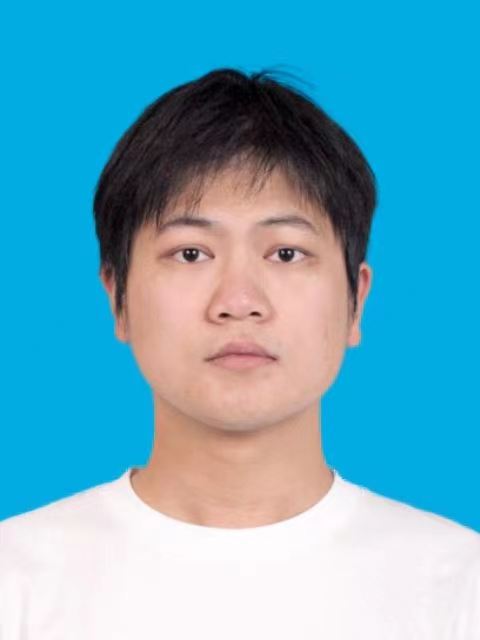}}]{Wei Zhang}
 received the B.E. degree in Automation from University of Science and Technology
of China (USTC), Hefei, China, in 2017 and is
pursuing the Ph.D. degree in Information and
Communication Engineering from University of
Science and Technology of China (USTC), Hefei,
China. His research interests include machine
learning, human-computer interaction and mixed
reality.
\end{IEEEbiography}
 \vspace{-10mm}
 \begin{IEEEbiography}[{\includegraphics[width=1in,height=1.25in,clip,keepaspectratio]{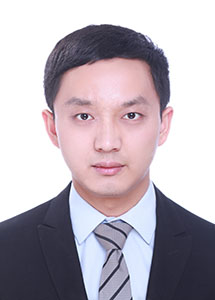}}]{Qinglang Guo}
received the PHD degree in Electronics and Information Engineering from University of Science and Technology of China, Hefei, China. And he works in China Academic of Electronics and Information Technology as an engineer since 2018. His research interests include large language model, knowledge graph and machine learning.
\end{IEEEbiography}
 \vspace{-10mm}

\begin{IEEEbiography}[{\includegraphics[width=1in,height=1.25in,clip,keepaspectratio]{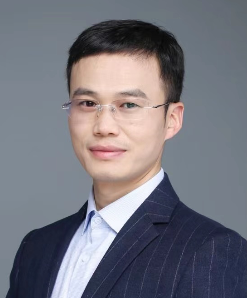}}]
		{Zhangjie Fu} (Member, IEEE) received the PhD degree in computer science from the College of Computer, Hunan University, China, in 2012. He is currently a professor with the School of Computer, Nanjing University of Information Science and Technology, China. His research interests include cloud and outsourcing security, digital forensics, network and information security. His research has been supported by NSFC, PAPD, and GYHY. He is a member of ACM.
	\end{IEEEbiography}
  \vspace{-10mm}
 \begin{IEEEbiography}[{\includegraphics[width=1in,height=1.25in,clip,keepaspectratio]{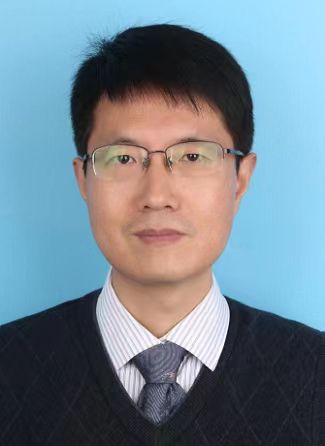}}]{Bin Li}
received the B.S. degree from the
Hefei University of Technology (HFUT), China,
in 1992, the M.S. degree from the Institute of
Plasma Physics, Chinese Academy of Sciences,
Hefei, China, in 1995, and the Ph.D degree
from the University of Science and Technology
of China (USTC), Hefei, China, in 2001. He is
currently a Professor with the School of Information Science and Technology, USTC. He has
authored or co-authored over 40 refereed publications. His current research interests include
evolutionary computation, pattern recognition, and human-computer interaction. 
\end{IEEEbiography}

\clearpage
\onecolumn
\appendix
\section{Supplementary Materials}
This supplementary material is organized as follows. We first provide the theoretical analyze of the proposed PFS in terms of active learning in Sec. \ref{sup:activate_learning} and the proof the Theorem 1 in Sec. \ref{sec:proof}. Additionally, we discuss the attack success rate and benign accuracy under poisoning rates with  $1.5 \%$ and  $2 \%$ on the CIFAR-10 dataset in Sec. \ref{sup:secA} and Sec. \ref{sup:secB}, respectively. Besides, we present the results of attack success rate under additional target classes on the CIFAR-10 dataset in Sec. \ref{sup:secC}. Finally, Sec. \ref{exp_cifar100} presents the experimental results on the CIFAR-100 dataset.

\subsection{Perspective of Active Learning}
\label{sup:activate_learning}
The process of selecting efficient poisoning samples in the context of backdoor attacks can be analogized to the principles of active learning. Active learning (AL) is a machine learning paradigm that efficiently acquires annotated data from an extensive pool of unlabeled data \cite{margatina2021active}. Its objective is to prioritize human labeling efforts on the most informative data points, leading to improved model performance while reducing data annotation costs. Similarly, the task of selecting efficient poisoning samples in a backdoor attack aims to identify the most informative data points that can optimize the efficiency of the poisoning process, thereby minimizing the costs for the attacker. A pivotal aspect of active learning encompasses the concepts of uncertainty and diversity in the dataset \cite{dasgupta2011two}. Methods that rely on uncertainty utilize the predictive confidence of model to select challenging examples, while diversity-based sampling leverages the dissimilarity among data points, often involving clustering techniques. Notably, the combination of uncertainty and diversity strategies has demonstrated significant success in active learning \cite{yuan2020cold,ru2020active}. Similarly, our proposed PFS takes into account both uncertainty (\textbf{Line 8} in Algorithm 1) and diversity (\textbf{Line 9} in Algorithm 1) of data points.

\begin{figure}[t!]
\setlength{\abovecaptionskip}{0.1cm}
    \setlength{\belowcaptionskip}{-0.3cm}
    \vspace{-1em}
    \centering
    \includegraphics[width=0.45\textwidth]{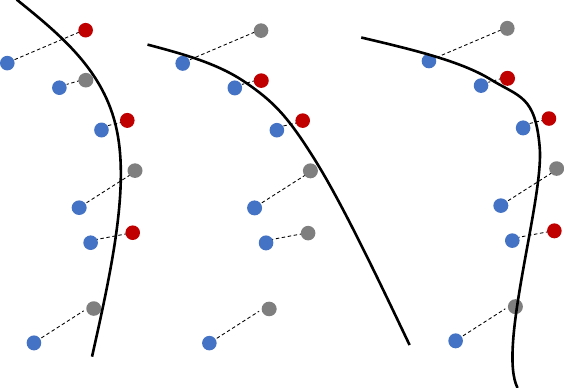}
    \caption{Demonstration of the balance between individual similarity and set diversity in formulating a poisoning set. The solid line is the boundary that separates benign and corresponding poisoning samples, allocated to different classes. Original benign samples are denoted by blue dots, while gray or red dots, connected to the blue ones, symbolize corresponding poisoning samples. Red dots represent the selected samples during data poisoning, with gray dots signifying non-selected ones. \textbf{Left:} The usage of a random sampling approach primarily emphasizes maximizing diversity. However, due to the lack of consideration for similarity, this results in non-compact boundaries. \textbf{Middle:} The approach is centered on similarity-based sampling, which generates a locally compact boundary around the red dots. Nevertheless, the clustering tendency of high-similarity points diminishes overall performance across the complete sample space compared to random sampling. \textbf{Right:} Accomplishing the balance  between individual similarity and set diversity, the approach yields a compact boundary while preserving representativeness throughout the entire sample space.}
    \label{Intuitively_analyses}
\end{figure}

Furthering our understanding of data point uncertainty within backdoor attacks, we delve into more comprehensive analyses to underscore the connection between the proposed measure of similarity between benign and corresponding poisoning samples and the expression of data point uncertainty. Drawing from theoretical insights provided by margin theory in active learning \cite{balcan2007margin}, we recognize that examples near the decision boundary offer significant potential for reducing annotation requirements. In the context of a backdoor attack, we exclusively consider the true class $y_i$ and the attack-target class $k$ of the clean sample $x_i$ and its corresponding poisoning sample $x'_i$, respectively, forming a two-class classification task. For any given poisoning sample $x'_i$, its corresponding clean sample $x_i$ is the nearest sample, albeit labeled with a different category. This configuration establishes a margin $M=\text{Dis}(x_i,x'_i)$, with the decision boundary residing centrally within the margin. As such, the distance between the sample $x_i$ and the decision boundary is proportionate to $\text{Dis}(x_i,x'_i)$. To summarize, the distance between benign and corresponding poisoning samples, $\text{Dis}(x_i,x'_i)$, can be employed to convey the uncertainty of data points. This assessment is carried out using a pre-trained feature extractor $E$ to compute distances within the feature space. Figure \ref{Intuitively_analyses} illustrates the effective integration of individual similarity and set diversity in constructing a poisoning injection.

\subsection{The Proof of Theorem 1}
\label{sec:proof}

\begin{theorem}
Suppose the training dataset consists of $N$ benign samples $\{(x_i,y_i)\}^{N}_{i=1}$ and $P$ poisoned samples $\{(x'_i,k)\}^{P}_{i=1}$, whose images are i.i.d. sampled from uniform distribution and belonging to $m$ classes. Assume that the DNN $f_\theta(\cdot)$ is a multivariate kernel regression $K(\cdot)$ and is trained via $\underset{\theta}\min \quad  \sum_{i=1}^{N} L\left(f_\theta(x), y\right)+
 \sum_{i=1}^{P} L\left(f_\theta\left(x^{\prime}\right), k\right)$. For the expected predictive confidences over the target label $k$, we have: $\mathbb{E}_{{x'_t}}\left[f_\theta({x'_t})\right]\propto \sum_{i=1}^{P} \frac{1}{({x}'_{{i}}-{x}_{{i}})\cdot ({x}'_{{i}}-{x}'_{{t}})}, i=1,\cdots,P $, where ${x'_t}$ is poisoned testing samples of attacks.
\end{theorem}

\begin{refproof}[Proof of Theorem 1:]
We treat our model as a m-way kernel least square classifier and use a cross-entropy loss for training the kernel. The output of $f_\theta(\cdot)$ is a m-dimensional vector. Let us assume $\phi_t(\cdot)\in \mathbb{R}$ be expected predictive confidences corresponding to the target class $k$. Following previous works \cite{guo2021aeva,guo2023scale}, we know the kernel regression solution for neural tangent kernel (NTK) \cite{jacot2018neural} is:
\begin{equation}
\phi_t(\cdot)=\frac{\sum_{i=1}^{N} K\left(\cdot, \boldsymbol{x}_{\boldsymbol{i}}\right) \cdot y_i+\sum_{i=1}^{P} K\left(\cdot, \boldsymbol{x}_{\boldsymbol{i}}^{\prime}\right) \cdot k}{\sum_{i=1}^{N} K\left(\cdot, \boldsymbol{x}_{\boldsymbol{i}}\right)+\sum_{i=1}^{P} K\left(\cdot, \boldsymbol{x}_{\boldsymbol{i}}^{\prime}\right)},
\end{equation}
where $K\left(\cdot, \cdot\right)$ is the RBF kernel and $K\left(\boldsymbol{x}, \boldsymbol{x}_{\boldsymbol{i}}\right)=e^{-2\gamma||\boldsymbol{x}-\boldsymbol{x}_{\boldsymbol{i}}||^2} (\gamma>0)$. Since the training samples are evenly distributed, there are $\frac{N}{m}$ benign samples belonging to $k$. Without loss of generality, we assume the target label $k=1$ while others are $0$. Then the regression solution can be re-formulated to:
\begin{equation}
\phi_t(\cdot)=\frac{\sum_{i=1}^{N/k} K\left(\cdot, \boldsymbol{x}_{\boldsymbol{i}}\right)+\sum_{i=1}^{P} K\left(\cdot, \boldsymbol{x}_{\boldsymbol{i}}^{\prime}\right)}{\sum_{i=1}^{N} K\left(\cdot, \boldsymbol{x}_{\boldsymbol{i}}\right)+\sum_{i=1}^{P} K\left(\cdot, \boldsymbol{x}_{\boldsymbol{i}}^{\prime}\right)},
\end{equation}
Therefore, following \cite{li2023towards}, we have:
\begin{equation}
\mathbb{E}_{{\boldsymbol{x}'_{\boldsymbol{i}}}}\left[f_\theta({\boldsymbol{x}'_{\boldsymbol{i}}})\right]\triangleq\phi_t(\boldsymbol{x}'_{\boldsymbol{t}})=\frac{\sum_{i=1}^{N/k} K\left(\boldsymbol{x}'_{\boldsymbol{t}}, \boldsymbol{x}_{\boldsymbol{i}}\right)+\sum_{i=1}^{P} K\left(\boldsymbol{x}'_{\boldsymbol{t}}, \boldsymbol{x}_{\boldsymbol{i}}^{\prime}\right)}{\sum_{i=1}^{N} K\left(\boldsymbol{x}'_{\boldsymbol{t}}, \boldsymbol{x}_{\boldsymbol{i}}\right)+\sum_{i=1}^{P} K\left(\boldsymbol{x}'_{\boldsymbol{t}}, \boldsymbol{x}_{\boldsymbol{i}}^{\prime}\right)}\approx\frac{\sum_{i=1}^{P} K\left(\boldsymbol{x}'_{\boldsymbol{t}}, \boldsymbol{x}_{\boldsymbol{i}}^{\prime}\right)}{\sum_{i=1}^{N} K\left(\boldsymbol{x}'_{\boldsymbol{t}}, \boldsymbol{x}_{\boldsymbol{i}}\right)+\sum_{i=1}^{P} K\left(\boldsymbol{x}'_{\boldsymbol{t}}, \boldsymbol{x}_{\boldsymbol{i}}^{\prime}\right)}.
\end{equation}
Similar to \cite{guo2023scale}, we also remove the term $\sum_{i=1}^{N/k} K\left(\boldsymbol{x}'_{\boldsymbol{t}}, \boldsymbol{x}_{\boldsymbol{i}}\right)$ because $\boldsymbol{x}'_{\boldsymbol{t}}$ typically don’t belong to the target $k$ and $\sum_{i=1}^{N/k} K\left(\boldsymbol{x}'_{\boldsymbol{t}}, \boldsymbol{x}_{\boldsymbol{i}}\right)  \ll \sum_{i=1}^{P} K\left(\boldsymbol{x}'_{\boldsymbol{t}}, \boldsymbol{x}'_{\boldsymbol{i}}\right)$, otherwise the attacker has no incentive to craft poisoned samples.

When $P$ close to $N$, which implies that the poisoning rate close to $50\%$, the attacker can achieve the optimal attack efficacy \cite{gu2019badnets,li2021invisible}. Given $N=P$, we have:
\begin{equation}
\begin{aligned}
    \phi_t(\boldsymbol{x}'_{\boldsymbol{t}})&=\sum_{i=1}^{P}\frac{ K\left(\boldsymbol{x}'_{\boldsymbol{t}}, \boldsymbol{x}_{\boldsymbol{i}}^{\prime}\right)}{ K\left(\boldsymbol{x}'_{\boldsymbol{t}}, \boldsymbol{x}_{\boldsymbol{i}}\right)+ K\left(\boldsymbol{x}'_{\boldsymbol{t}}, \boldsymbol{x}_{\boldsymbol{i}}^{\prime}\right)}=\sum_{i=1}^{P}\frac{1}{1+\frac{K\left(\boldsymbol{x}'_{\boldsymbol{t}}, \boldsymbol{x}_{\boldsymbol{i}}\right)}{K\left(\boldsymbol{x}'_{\boldsymbol{t}}, \boldsymbol{x}_{\boldsymbol{i}}^{\prime}\right)}}=\sum_{i=1}^{P}\frac{1}{1+\frac{e^{-2\gamma||\boldsymbol{x}_{\boldsymbol{t}}^{\prime}-\boldsymbol{x}_{\boldsymbol{i}}||^2}}{e^{-2\gamma||\boldsymbol{x}_{\boldsymbol{t}}^{\prime}-\boldsymbol{x}'_{\boldsymbol{i}}||^2}}}\\
    &=\sum_{i=1}^{P}\frac{1}{1+e^{ 2\gamma||\boldsymbol{x}_{\boldsymbol{t}}^{\prime}-\boldsymbol{x}'_{\boldsymbol{i}}||^2-2\gamma||\boldsymbol{x}_{\boldsymbol{t}}^{\prime}-\boldsymbol{x}_{\boldsymbol{i}}||^2}}=\sum_{i=1}^{P}\frac{1}{1+e^{ 2\gamma(\boldsymbol{x}'_{\boldsymbol{i}}-\boldsymbol{x}_{\boldsymbol{i}})\cdot (\boldsymbol{x}'_{\boldsymbol{i}}+\boldsymbol{x}_{\boldsymbol{i}}-2\boldsymbol{x}'_{\boldsymbol{t}})}}\\
    &\approx \sum_{i=1}^{P}\frac{1}{1+e^{ 4\gamma(\boldsymbol{x}'_{\boldsymbol{i}}-\boldsymbol{x}_{\boldsymbol{i}})\cdot (\boldsymbol{x}'_{\boldsymbol{i}}-\boldsymbol{x}'_{\boldsymbol{t}})}}
    \end{aligned}
\end{equation}

we here let $\boldsymbol{x}_{\boldsymbol{i}}=\boldsymbol{x}'_{\boldsymbol{i}}$ in $\boldsymbol{x}'_{\boldsymbol{i}}+\boldsymbol{x}_{\boldsymbol{i}}-2\boldsymbol{x}'_{\boldsymbol{t}}$, because $\boldsymbol{x}'_{\boldsymbol{i}}$ typically be similar as $\boldsymbol{x}_{\boldsymbol{i}}$ and $\boldsymbol{x}'_{\boldsymbol{i}}-\boldsymbol{x}_{\boldsymbol{i}}\ll \boldsymbol{x}'_{\boldsymbol{i}}-\boldsymbol{x}'_{\boldsymbol{t}}$. Therefore, we have:

\begin{equation}
    \mathbb{E}_{{x'_t}}\left[f_\theta({x'_t})\right]\propto \sum_{i=1}^{P} \frac{1}{(\boldsymbol{x}'_{\boldsymbol{i}}-\boldsymbol{x}_{\boldsymbol{i}})\cdot (\boldsymbol{x}'_{\boldsymbol{i}}-\boldsymbol{x}'_{\boldsymbol{t}})}.
\end{equation}

\end{refproof}

\subsection{Attack Success Rate Under Additional poisoning rates on the CIFAR-10 Dataset}
\label{sup:secA}
We evaluate the attack success rate (ASR) under different poisoning rates on the CIFAR-10 dataset. Tables \ref{tab:cifar10_1} and \ref{tab:cifar10_2} present the ASR with trigger BadNets and Blended at poisoning rates of $1.5 \%$ and $2 \%$, respectively. Results under poisoning rates of $1.5 \%$ and $2 \%$ are similar to those at $1 \%$. 

\textbf{(i) Experiments under poisoning rates of $1.5 \%$:} Our PFS approach outperforms random selection in all cases, with average boosts of 0.030 and 0.055 on BadNets and Blended triggers, respectively. The PFS method also achieves better attack results than FUS in most cases (27/36). The combination of these two methods achieves superior results in all 36 settings.

\textbf{(ii) Experiments under poisoning rates of $2 \%$:} Our PFS approach outperforms random selection in all cases, with average boosts of 0.023 and 0.04 on BadNets and Blended triggers, respectively. The PFS method also achieves better attack results than FUS in most cases (26/36). The combination of these two methods achieves superior results in all 36 settings.

\begin{table*}
	\caption{The attack success rate (ASR) on the CIFAR-10 dataset. All results are averaged over 5 different runs. The poisoning rates of experiments with trigger BadNets and Blended are $1.5 \%$, respectively. Among four select strategies, the best result is denoted in \textbf{boldface} while the \underline{underline} indicates the second-best result. The settings corresponding to the gray background are represented as the proxy poisoning attack within the attacker phase and the actual poisoning process within the victim phase being consistent in FUS.
	\label{tab:cifar10_1}}
	\centering
	\scriptsize
	\scalebox{1}{
		\begin{tabular}{cccccc|cccc|cccc}
			\toprule
			\multirow{3}{*}{\makecell*[c]{Trigger}} &\multirow{3}{*}{Data Transform} &\multicolumn{12}{c}{\makecell*[c]{Model}} \\ 
			\cline{3-14} 
			\specialrule{0em}{1pt}{1pt}
			&&
			\multicolumn{4}{c}{\makecell*[c]{VGG16}}&\multicolumn{4}{c}{\makecell*[c]{ResNet18}}&\multicolumn{4}{c}{\makecell*[c]{PreAct-ResNet18}}\\
			
			&&Random&PFS&FUS&FUS+PFS&Random&PFS&FUS &FUS+PFS&Random&PFS&FUS&FUS+PFS\\
			\midrule
			\multirow{7}{*}{BadNets}&None&0.961&\underline{0.986}&0.982&\textbf{0.992}&0.976&\underline{0.993}&0.990&\textbf{0.997}&0.973&\underline{0.992}&0.989&\textbf{0.997}\\
			&RandomCrop&0.924&\underline{0.970}&0.966&\textbf{0.983}&0.944&\underline{0.980}&0.974&\textbf{0.988}&0.946&\underline{0.979}&0.977&\textbf{0.989}\\
			&RandomHorizontalFlip&0.952&\underline{0.983}&0.979&\textbf{0.991}&0.973&0.975&\underline{0.986}&\textbf{0.995}&0.967&\underline{0.988}&0.984&\textbf{0.994}\\
			&RandomRotation&0.915&\underline{0.968}&0.958&\textbf{0.977}&0.929&\underline{0.979}&0.971&\textbf{0.988}&0.941&\underline{0.979}&0.973&\textbf{0.988}\\
			&ColorJitter&0.958&\underline{0.986}&0.981&\textbf{0.992}&0.977&\underline{0.994}&0.990&\textbf{0.997}&0.972&\underline{0.993}&0.988&\textbf{0.996}\\
			&{\makecell[c]{RandomCrop+ \\ RandomHorizontalFlip}}&0.932&\underline{0.970}&\cellcolor{mygray}0.965&\cellcolor{mygray}\textbf{0.981}&0.949&\underline{0.980}&0.978&\textbf{0.989}&0.952&\underline{0.981}&0.978&\textbf{0.989}\\
			\cline{2-14}
			\specialrule{0em}{1pt}{1pt}
			&Avg&0.940&\underline{0.977}&0.972&\textbf{0.986}&0.958&\underline{0.984}&0.982&\textbf{0.992}&0.959&\underline{0.985}&0.982&\textbf{0.992}\\
			\cmidrule{1-14} 
			\multirow{7}{*}{Blended}&None&0.866&\underline{0.921}&0.918&\textbf{0.953}&0.883&\underline{0.954}&0.939&\textbf{0.971}&0.909&\underline{0.970}&0.954&\textbf{0.982}\\
			&RandomCrop&0.855&0.900&\underline{0.933}&\textbf{0.942}&0.879&0.935&\underline{0.943}&\textbf{0.963}&0.882&\underline{0.948}&0.938&\textbf{0.967}\\
			&RandomHorizontalFlip&0.889&0.937&\underline{0.952}&\textbf{0.970}&0.906&\underline{0.966}&0.960&\textbf{0.983}&0.927&\underline{0.975}&0.961&\textbf{0.987}\\
			&RandomRotation&0.859&0.909&\underline{0.929}&\textbf{0.937}&0.894&0.938&\underline{0.947}&\textbf{0.967}&0.901&\underline{0.955}&0.953&\textbf{0.972}\\
			&ColorJitter&0.851&\underline{0.915}&0.912&\textbf{0.938}&0.872&\underline{0.950}&0.929&\textbf{0.969}&0.913&\underline{0.966}&0.945&\textbf{0.976}\\
			&{\makecell[c]{RandomCrop+ \\ RandomHorizontalFlip}}&0.885&0.931&\cellcolor{mygray}\underline{0.952}&\cellcolor{mygray}\textbf{0.965}&0.905&0.958&\underline{0.964}&\textbf{0.982}&0.919&\underline{0.967}&\underline{0.967}&\textbf{0.984}\\
			\cline{2-14}
			\specialrule{0em}{1pt}{1pt}
			&Avg&0.868&0.919&\underline{0.933}&\textbf{0.951}&0.890&\underline{0.950}&0.947&\textbf{0.973}&0.909&\underline{0.964}&0.953&\textbf{0.978}\\
			\bottomrule
	\end{tabular}}
\end{table*}

\begin{table*}
	\caption{The attack success rate (ASR) on the CIFAR-10 dataset. All results are averaged over 5 different runs. The poisoning rates of experiments with trigger BadNets and Blended are $2 \%$, respectively. Among four select strategies, the best result is denoted in \textbf{boldface} while the \underline{underline} indicates the second-best result. The settings corresponding to the gray background are represented as the proxy poisoning attack within the attacker phase and the actual poisoning process within the victim phase being consistent in FUS.
	\label{tab:cifar10_2}}
	\centering
	\scriptsize
	\scalebox{1}{
		\begin{tabular}{cccccc|cccc|cccc}
			\toprule
			\multirow{3}{*}{\makecell*[c]{Trigger}} &\multirow{3}{*}{Data Transform} &\multicolumn{12}{c}{\makecell*[c]{Model}} \\ 
			\cline{3-14} 
			\specialrule{0em}{1pt}{1pt}
			&&
			\multicolumn{4}{c}{\makecell*[c]{VGG16}}&\multicolumn{4}{c}{\makecell*[c]{ResNet18}}&\multicolumn{4}{c}{\makecell*[c]{PreAct-ResNet18}}\\
			
			&&Random&PFS&FUS&FUS+PFS&Random&PFS&FUS &FUS+PFS&Random&PFS&FUS&FUS+PFS\\
			\midrule
			\multirow{7}{*}{BadNets}&None&0.971&\underline{0.987}&0.986&\textbf{0.993}&0.982&\underline{0.994}&0.992&\textbf{0.998}&0.977&\underline{0.993}&0.991&\textbf{0.997}\\
			&RandomCrop&0.940&\underline{0.973}&0.970&\textbf{0.986}&0.954&\underline{0.980}&0.979&\textbf{0.991}&0.946&\underline{0.982}&0.979&\textbf{0.992}\\
			&RandomHorizontalFlip&0.961&\underline{0.984}&0.981&\textbf{0.992}&0.976&\underline{0.991}&0.988&\textbf{0.995}&0.975&\underline{0.989}&0.986&\textbf{0.995}\\
			&RandomRotation&0.931&\underline{0.969}&0.966&\textbf{0.986}&0.948&\underline{0.977}&\underline{0.977}&\textbf{0.992}&0.946&\underline{0.981}&0.979&\textbf{0.991}\\
			&ColorJitter&0.966&\underline{0.986}&0.984&\textbf{0.993}&0.982&\underline{0.993}&0.992&\textbf{0.998}&0.981&\underline{0.992}&0.990&\textbf{0.997}\\
			&{\makecell[c]{RandomCrop+ \\ RandomHorizontalFlip}}&0.938&\underline{0.974}&\cellcolor{mygray}0.972&\cellcolor{mygray}\textbf{0.986}&0.957&\underline{0.982}&0.980&\textbf{0.992}&0.959&\underline{0.984}&0.981&\textbf{0.991}\\
		    \cline{2-14}
		    \specialrule{0em}{1pt}{1pt}
			&Avg&0.951&\underline{0.979}&0.977&\textbf{0.989}&0.967&\underline{0.986}&0.985&\textbf{0.994}&0.964&\underline{0.987}&0.984&\textbf{0.994}\\
			\cmidrule{1-14} 
			\multirow{7}{*}{Blended}&None&0.914&\underline{0.943}&\textbf{0.952}&\textbf{0.952}&0.916&\underline{0.969}&0.955&\textbf{0.981}&0.937&\underline{0.978}&0.967&\textbf{0.988}\\
			&RandomCrop&0.896&0.937&\underline{0.958}&\textbf{0.966}&0.912&\underline{0.961}&\underline{0.961}&\textbf{0.978}&0.913&0.960&\underline{0.965}&\textbf{0.985}\\
			&RandomHorizontalFlip&0.923&0.955&\underline{0.972}&\textbf{0.983}&0.931&\underline{0.976}&0.973&\textbf{0.990}&0.947&\underline{0.982}&0.976&\textbf{0.991}\\
			&RandomRotation&0.900&0.936&\underline{0.952}&\textbf{0.963}&0.915&\underline{0.958}&\underline{0.958}&\textbf{0.973}&0.922&\underline{0.965}&\underline{0.965}&\textbf{0.982}\\
			&ColorJitter&0.904&0.940&\underline{0.951}&\textbf{0.963}&0.918&\underline{0.965}&0.949&\textbf{0.979}&0.938&\underline{0.973}&0.964&\textbf{0.986}\\
			&{\makecell[c]{RandomCrop+ \\ RandomHorizontalFlip}}&0.922&0.955&\cellcolor{mygray}\underline{0.973}&\cellcolor{mygray}\textbf{0.986}&0.927&0.968&\underline{0.977}&\textbf{0.991}&0.933&0.973&\underline{0.979}&\textbf{0.992}\\
			\cline{2-14}
			\specialrule{0em}{1pt}{1pt}
			&Avg&0.910&0.944&\underline{0.960}&\textbf{0.969}&0.920&\underline{0.966}&0.962&\textbf{0.969}&0.932&\underline{0.972}&0.969&\textbf{0.987}\\
			\bottomrule
	\end{tabular}}
\end{table*}
\subsection{Benign Accuracy Under Additional poisoning rates on the CIFAR-10 Dataset}
\label{sup:secB}
We also present the benign accuracy (BA) (\textit{i.e.}, the probability of classifying a benign test data to the correct label) under other poisoning rates on the CIFAR-10 dataset. Table~\ref{tab:cifar10_3} and Table~\ref{tab:cifar10_4} illustrate the benign accuracy with trigger BadNets and Blended under poisoning rates of $1.5 \%$ and $2 \%$, respectively. The results show the similar BA with four strategies, which demonstrates that our proposed strategy has no negative impact on benign accuracy. 

\begin{table*}
	\caption{The benign accuracy (BA) on the CIFAR-10 dataset. All results are averaged over 5 different runs. The poisoning rates of experiments with trigger BadNets and Blended are $1.5 \%$, respectively.
	\label{tab:cifar10_3}}
	\centering
	\scriptsize
	\scalebox{1}{
		\begin{tabular}{cccccc|cccc|cccc}
			\toprule
			\multirow{3}{*}{\makecell*[c]{Trigger}} &\multirow{3}{*}{Data Transform} &\multicolumn{12}{c}{\makecell*[c]{Model}} \\ 
			\cline{3-14} 
			\specialrule{0em}{1pt}{1pt}
			&&
			\multicolumn{4}{c}{\makecell*[c]{VGG16}}&\multicolumn{4}{c}{\makecell*[c]{ResNet18}}&\multicolumn{4}{c}{\makecell*[c]{PreAct-ResNet18}}\\
			
			&&Random&PFS&FUS&FUS+PFS&Random&PFS&FUS &FUS+PFS&Random&PFS&FUS&FUS+PFS\\
			\midrule
			\multirow{7}{*}{BadNets}&None&0.868&0.868&0.865&0.866&0.837&0.837&0.838&0.835&0.850&0.849&0.848&0.847\\
			&RandomCrop&0.903&0.905&0.905&0.903&0.914&0.912&0.914&0.911&0.914&0.914&0.914&0.914\\
			&RandomHorizontalFlip&0.892&0.892&0.893&0.894&0.878&0.879&0.879&0.878&0.886&0.888&0.886&0.883\\
			&RandomRotation&0.876&0.873&0.876&0.874&0.883&0.881&0.880&0.880&0.889&0.888&0.889&0.889\\
			&ColorJitter&0.860&0.860&0.860&0.859&0.833&0.829&0.833&0.830&0.843&0.844&0.841&0.843\\
			&{\makecell[c]{RandomCrop+ \\ RandomHorizontalFlip}}&0.920&0.919&0.920&0.921&0.927&0.926&0.928&0.925&0.927&0.928&0.929&0.927\\
			\cline{2-14}
			\specialrule{0em}{1pt}{1pt}
			&Avg&0.887&0.886&0.887&0.886&0.879&0.877&0.879&0.877&0.885&0.885&0.885&0.884\\
			\cmidrule{1-14} 
			\multirow{7}{*}{Blended}&None&0.865&0.866&0.865&0.866&0.836&0.832&0.832&0.833&0.846&0.846&0.849&0.847\\
			&RandomCrop&0.904&0.906&0.904&0.904&0.912&0.913&0.912&0.913&0.913&0.914&0.915&0.914\\
			&RandomHorizontalFlip&0.892&0.892&0.889&0.891&0.876&0.878&0.875&0.875&0.882&0.884&0.885&0.887\\
			&RandomRotation&0.873&0.875&0.875&0.873&0.880&0.880&0.881&0.880&0.887&0.888&0.889&0.889\\
			&ColorJitter&0.858&0.859&0.860&0.858&0.829&0.832&0.824&0.828&0.839&0.841&0.843&0.842\\
			&{\makecell[c]{RandomCrop+ \\ RandomHorizontalFlip}}&0.922&0.920&0.919&0.920&0.928&0.927&0.928&0.928&0.928&0.927&0.929&0.928\\
			\cline{2-14}
			\specialrule{0em}{1pt}{1pt}
			&Avg&0.886&0.886&0.885&0.885&0.877&0.877&0.875&0.876&0.883&0.883&0.885&0.885\\
			\bottomrule
	\end{tabular}}
\end{table*}

\begin{table*}
	\caption{The benign accuracy (BA) on the CIFAR-10 dataset. All results are averaged over 5 different runs. The poisoning rates of experiments with trigger BadNets and Blended are $2 \%$, respectively.
	\label{tab:cifar10_4}}
	\centering
	\scriptsize
	\scalebox{1}{
		\begin{tabular}{cccccc|cccc|cccc}
			\toprule
			\multirow{3}{*}{\makecell*[c]{Trigger}} &\multirow{3}{*}{Data Transform} &\multicolumn{12}{c}{\makecell*[c]{Model}} \\ 
			\cline{3-14} 
			\specialrule{0em}{1pt}{1pt}
			&&
			\multicolumn{4}{c}{\makecell*[c]{VGG16}}&\multicolumn{4}{c}{\makecell*[c]{ResNet18}}&\multicolumn{4}{c}{\makecell*[c]{PreAct-ResNet18}}\\
			
			&&Random&PFS&FUS&FUS+PFS&Random&PFS&FUS &FUS+PFS&Random&PFS&FUS&FUS+PFS\\
			\midrule
			\multirow{7}{*}{BadNets}&None&0.867&0.865&0.866&0.863& 0.837&0.835&0.835&0.833
			&0.851&0.846&0.848&0.847\\
			&RandomCrop&0.904&0.905&0.905&0.905
			&0.913&0.914&0.912&0.912
			&0.916&0.915&0.913&0.913\\
			&RandomHorizontalFlip&0.891&0.891&0.891&0.891
			&0.878&0.878&0.878&0.876
			&0.886&0.885&0.886&0.885\\
			&RandomRotation&0.873&0.876&0.874&0.872
			&0.882&0.882&0.881&0.880
			&0.888&0.889&0.889&0.890\\
			&ColorJitter&0.862&0.857&0.857&0.859
			&0.830&0.827&0.829&0.831
			&0.844&0.843&0.843&0.841\\
			&{\makecell[c]{RandomCrop+ \\ RandomHorizontalFlip}}&0.920&0.920&0.919&0.919
			&0.928&0.926&0.926&0.925
			&0.928&0.930&0.928&0.927\\
			\cline{2-14}
			\specialrule{0em}{1pt}{1pt}
			&Avg&0.886&0.886&0.885&0.885&0.878&0.877&0.877&0.876&0.886&0.885&0.885&0.884\\
			\cmidrule{1-14} 
			\multirow{7}{*}{Blended}&None&0.865&0.866&0.865&0.866&0.836&0.832&0.832&0.833&0.846&0.846&0.849&0.847\\
			&RandomCrop&0.904&0.906&0.904&0.904&0.912&0.913&0.912&0.913&0.913&0.914&0.915&0.914\\
			&RandomHorizontalFlip&0.892&0.892&0.889&0.891&0.876&0.878&0.875&0.875&0.882&0.884&0.885&0.887\\
			&RandomRotation&0.873&0.875&0.875&0.873&0.880&0.880&0.881&0.880&0.887&0.888&0.889&0.889\\
			&ColorJitter&0.858&0.859&0.860&0.858&0.829&0.832&0.824&0.828&0.839&0.841&0.843&0.842\\
			&{\makecell[c]{RandomCrop+ \\ RandomHorizontalFlip}}&0.922&0.920&0.919&0.920&0.928&0.927&0.928&0.928&0.928&0.927&0.929&0.928\\
			\cline{2-14}
			\specialrule{0em}{1pt}{1pt}
			&Avg&0.886&0.886&0.885&0.885&0.877&0.877&0.875&0.876&0.883&0.883&0.885&0.885\\
			\bottomrule
	\end{tabular}}
\end{table*}

\subsection{Attack Success Rate Under Additional Target Classes on the CIFAR-10 Dataset}
\label{sup:secC}
This section presents the attack success rate (ASR) under additional target classes on the CIFAR-10 dataset. Table~\ref{tab:cifar10_5} illustrates the attack success rate with trigger BadNets and Blended under target category 3. Compared with random selection, our PFS approach outperforms in all cases, with average boosts of 0.045 and 0.076 on BadNets and Blended triggers. The PFS method also achieves better attack results in most cases compared to FUS (22/24). The combination of these two achieves superior results in all 24 settings. These outcomes consistently confirm the conclusions we reached in the paper.

\begin{table*}
	\caption{The attack success rate (ASR) on the CIFAR-10 dataset. In this experiment, attack target k is set to category 3. All results are averaged over 5 different runs. The poisoning rates of experiments with trigger BadNets and Blended are $1 \%$, respectively. Among four select strategies, the best result is denoted in \textbf{boldface} while the \underline{underline} indicates the second-best result. The settings corresponding to the gray background are represented as the proxy poisoning attack within the attacker phase and the actual poisoning process within the victim phase being consistent in FUS.
	\label{tab:cifar10_5}}
	\centering
	\small
	\scalebox{1}{
		\begin{tabular}{cccccc|cccc}
			\toprule
			\multirow{3}{*}{\makecell*[c]{Trigger}} &\multirow{3}{*}{Data Transform} &\multicolumn{8}{c}{\makecell*[c]{Model}} \\ 
			\cline{3-10} 
			\specialrule{0em}{1pt}{1pt}
			&&
			\multicolumn{4}{c}{\makecell*[c]{VGG16}}&\multicolumn{4}{c}{\makecell*[c]{ResNet18}}\\
			
			&&Random&PFS&FUS&FUS+PFS&Random&PFS&FUS&FUS+PFS\\
			\midrule
			\multirow{7}{*}{BadNets}&None&0.950&\underline{0.984}&0.977&\textbf{0.986}&0.959&\underline{0.991}&0.985&\textbf{0.994}\\
			&RandomCrop&0.909&\underline{0.960}&0.954&\textbf{0.968}&0.925&\underline{0.976}&0.966&\textbf{0.979}\\
			&RandomHorizontalFlip&0.934&\underline{0.980}&0.972&\textbf{0.983}&0.959&\underline{0.987}&0.981&\textbf{0.990}\\
			&RandomRotation&0.890&\textbf{0.951}&\underline{0.942}&\textbf{0.951}&0.912&\underline{0.968}&0.955&\textbf{0.972}\\
			&ColorJitter&0.945&\underline{0.983}&0.975&\textbf{0.986}&0.966&\underline{0.991}&0.985&\textbf{0.993}\\
			&{\makecell[c]{RandomCrop+ \\ RandomHorizontalFlip}}&0.905&\underline{0.964}&\cellcolor{mygray}0.957&\cellcolor{mygray}\textbf{0.968}&0.925&\underline{0.978}&0.968&\textbf{0.980}\\
		    \cline{2-10}
		    \specialrule{0em}{1pt}{1pt}
			&Avg&0.922&\underline{0.970}&0.963&\textbf{0.974}&0.941&\underline{0.982}&0.973&\textbf{0.985}\\
			\cmidrule{1-10} 
			\multirow{7}{*}{Blended}&None&0.774&\underline{0.867}&0.851&\textbf{0.899}&0.806&\underline{0.915}&0.850&\textbf{0.935}\\
			&RandomCrop&0.780&0.813&\underline{0.822}&\textbf{0.869}&0.791&\underline{0.902}&0.827&\textbf{0.918}\\
			&RandomHorizontalFlip&0.836&\underline{0.899}&0.887&\textbf{0.934}&0.845&\underline{0.949}&0.888&\textbf{0.962}\\
			&RandomRotation&0.793&\underline{0.832}&0.813&\textbf{0.847}&0.804&\underline{0.898}&0.840&\textbf{0.903}\\
			&ColorJitter&0.778&\underline{0.836}&0.805&\textbf{0.874}&0.814&\underline{0.907}&0.854&\textbf{0.927}\\
			&{\makecell[c]{RandomCrop+ \\ RandomHorizontalFlip}}&0.815&0.843&\cellcolor{mygray}\underline{0.871}&\cellcolor{mygray}\textbf{0.884}&0.838&\underline{0.927}&0.893&\textbf{0.951}\\
			\cline{2-10}
			\specialrule{0em}{1pt}{1pt}
			&Avg&0.796&\underline{0.848}&0.842&\textbf{0.885}&0.816&\underline{0.916}&0.859&\textbf{0.933}\\
			\bottomrule
	\end{tabular}}
\end{table*}

\subsection{Experiments on CIFAR-100 Dataset}
\label{exp_cifar100}
The results for CIFAR-100 in Table~\ref{tab:cifar100} are similar to those for CIFAR-10. Our PFS approach outperforms random selection in all cases, with average boosts of 0.047, 0.068, and 0.037 for BadNets, Blended, and Optimized triggers, respectively. Compared to FUS, PFS achieves better attack results in 49 out of 54 settings. 

\begin{table*}
	\centering
	\scriptsize
 \caption{The attack success rate on CIFAR-100 dataset. All results are averaged over 5 different runs. The poisoning rates of experiments with trigger BadNets, Blended, and Optimized are $1.5\%$, $1.5\%$, and $0.75\%$, respectively. Among four select strategies, the best result is denoted in \textbf{boldface} while the \underline{underline} indicates the second-best result. The settings corresponding to the gray background are represented as the proxy poisoning attack within the attacker phase and the actual poisoning process within the victim phase being consistent in FUS.
	\label{tab:cifar100}}
	\scalebox{0.9}{
		\begin{tabular}{cccccc|cccc|cccc}
			\toprule
			\multirow{3}{*}{\makecell*[c]{Trigger}} &\multirow{3}{*}{Data Transform} &\multicolumn{12}{c}{\makecell*[c]{Model}} \\ 
			\cline{3-14} 
			\specialrule{0em}{1pt}{1pt}
			&&
			\multicolumn{4}{c}{\makecell*[c]{VGG16}}&\multicolumn{4}{c}{\makecell*[c]{ResNet18}}&\multicolumn{4}{c}{\makecell*[c]{PreAct-ResNet18}}\\
			
			&&Random&PFS&FUS&FUS+PFS&Random&PFS&FUS &FUS+PFS&Random&PFS&FUS&FUS+PFS\\
			\midrule
			\multirow{7}{*}{BadNets}&None&0.933&\underline{0.970}&0.965&\textbf{0.971}&0.950&\underline{0.978}&0.973&\textbf{0.979}&0.949&\textbf{0.977}&\underline{0.970}&\textbf{0.977}\\
			&RandomCrop&0.882&\textbf{0.947}&0.930&\underline{0.945}&0.894&\underline{0.951}&0.940&\textbf{0.952}&0.888&\textbf{0.951}&\underline{0.940}&\textbf{0.951}\\
			&RandomHorizontalFlip&0.920&\underline{0.964}&0.958&\textbf{0.965}&0.936&\underline{0.970}&0.965&\textbf{0.971}&0.929&\textbf{0.968}&\underline{0.962}&\textbf{0.968}\\
			&RandomRotation&0.868&\underline{0.923}&0.912&\textbf{0.924}&0.866&\textbf{0.939}&0.925&\underline{0.938}&0.875&\underline{0.935}&0.926&\textbf{0.938}\\
			&ColorJitter&0.934&\underline{0.970}&0.965&\textbf{0.971}&0.952&\underline{0.978}&0.973&\textbf{0.979}&0.946&\underline{0.977}&0.970&\textbf{0.978}\\
			&{\makecell[c]{RandomCrop+ \\ RandomHorizontalFlip}}&0.879&\textbf{0.939}&\cellcolor{mygray}\underline{0.929}&\cellcolor{mygray}\textbf{0.939} &0.888&\underline{0.947}&0.942&\textbf{0.950}&0.894&\underline{0.949}&0.941&\textbf{0.951}\\
			\cline{2-14}
			\specialrule{0em}{1pt}{1pt}
			&Avg&0.903&\underline{0.952}&0.943&\textbf{0.953}&0.914&\underline{0.961}&0.953&\textbf{0.962}&0.914&\underline{0.960}&0.952&\textbf{0.961}\\
			\cmidrule{1-14}
			\multirow{7}{*}{Blended}&None&0.866&\underline{0.923}&0.920&\textbf{0.941}&0.884&\underline{0.963}&0.946&\textbf{0.974}&0.894&\underline{0.975}&0.950&\textbf{0.982}\\
			&RandomCrop&0.863&0.874&\textbf{0.891}&\underline{0.880}&0.826&\underline{0.916}&0.906&\textbf{0.938}&0.829&\underline{0.931}&0.912&\textbf{0.942}\\
			&RandomHorizontalFlip&0.891&\underline{0.945}&0.935&\textbf{0.960}&0.902&\underline{0.968}&0.951&\textbf{0.977}&0.912&\underline{0.974}&0.958&\textbf{0.982}\\
			&RandomRotation&0.818&0.862&\textbf{0.901}&\underline{0.865}&0.841&\underline{0.916}&\underline{0.916}&\textbf{0.936}&0.848&\underline{0.926}&0.921&\textbf{0.944}\\
			&ColorJitter&0.863&\underline{0.920}&0.917&\textbf{0.931}&0.904&\underline{0.972}&0.956&\textbf{0.977}&0.913&\underline{0.979}&0.959&\textbf{0.983}\\
			&{\makecell[c]{RandomCrop+ \\ RandomHorizontalFlip}}&0.811&0.865&\cellcolor{mygray}\textbf{0.905}&\cellcolor{mygray}\underline{0.872}&0.826&0.918&\underline{0.920}&\textbf{0.942}&0.850&\underline{0.931}&0.924&\textbf{0.947}\\
			\cline{2-14}
			\specialrule{0em}{1pt}{1pt}
			&Avg&0.852&0.898&\textbf{0.912}&\underline{0.908}&0.864&\underline{0.942}&0.933&\textbf{0.957}&0.874&\underline{0.953}&0.937&\textbf{0.963}\\
			\cmidrule{1-14} 
			\multirow{7}{*}{Optimized}&None&0.979&\textbf{0.999}&\underline{0.995}&\textbf{0.999}&0.884&\underline{0.972}&0.931&\textbf{0.982}&0.895&\underline{0.983}&0.962&\textbf{0.987}\\
		    &RandomCrop&0.989&\underline{0.999}&0.997&\textbf{1.000}&0.983&\underline{0.999}&0.996&\textbf{1.000}&0.981&\underline{0.999}&0.995&\textbf{1.000}\\
			&RandomHorizontalFlip&0.985&\underline{0.999}&0.997&\textbf{1.000}&0.938&\textbf{0.998}&0.980&\underline{0.983}&0.938&\underline{0.997}&0.984&\textbf{0.998}\\
			&RandomRotation&0.810&\underline{0.831}&\textbf{0.867}&0.785&0.827&\textbf{0.909}&\underline{0.882}&\underline{0.882}&0.835&\textbf{0.908}&0.883&\underline{0.900}\\
			&ColorJitter&0.984&\underline{0.997}&0.993&\textbf{1.000}&0.967&\textbf{0.989}&0.979&\underline{0.985}&0.956&\underline{0.993}&0.990&\textbf{0.994}\\
			&{\makecell[c]{RandomCrop+ \\ RandomHorizontalFlip}}&0.991&\underline{0.999}&\cellcolor{mygray}0.997&\cellcolor{mygray}\textbf{1.000}&0.987&\underline{0.999}&0.998&\textbf{1.000}&0.982&\underline{0.998}&0.994&\textbf{0.999}\\
			\cline{2-14}
			\specialrule{0em}{1pt}{1pt}
			&Avg&0.956&\underline{0.971}&\textbf{0.974}&0.964&0.931&\textbf{0.978}&0.961&\underline{0.972}&0.931&\textbf{0.980}&\underline{0.968}&\textbf{0.980}\\
			\bottomrule
	\end{tabular}}
\end{table*}

\vfill

\end{document}